\documentclass[preprint]{aastex}

\usepackage{natbib}
\usepackage{amsmath} \usepackage{amssymb}

\shorttitle{Nonparametric Analysis of WMAP Power Spectra}
\shortauthors{Aghamousa, Arjunwadkar, and Souradeep}

\begin{document}

\title{Evolution of the CMB Power Spectrum Across WMAP Data Releases: A Nonparametric Analysis}

\author{Amir Aghamousa}
\affil{Centre for Modeling and Simulation, University of Pune, Pune 411 007 India}
\email{amir@cms.unipune.ac.in}

\author{Mihir Arjunwadkar}
\affil{Centre for Modeling and Simulation, University of Pune, Pune 411 007 India,}
\affil{National Centre for Radio Astrophysics, University of Pune Campus, Pune 411 007 India}
\email{mihir@ncra.tifr.res.in}

\and

\author{Tarun Souradeep}
\affil{Inter-University Centre for Astronomy and Astrophysics, University of Pune Campus, Pune 411 007 India}
\email{tarun@iucaa.ernet.in}

\begin{abstract}
 Using a nonparametric function estimation methodology,
 we present a comparative analysis of the WMAP 1-, 3-, 5-, and 7-year data releases
 for the CMB angular power spectrum with respect to the following key questions:
 (a) How well is the  power spectrum determined by the data alone?
 (b) How well is the $\Lambda$CDM model supported by a model-independent, data-driven analysis?
 (c) What are the realistic uncertainties on peak/dip locations and heights?
 Our results show that the height of the power spectrum is well determined by data alone for
 multipole $l$ approximately less than 546 (1-year), 667 (3-year), 804 (5-year), and 842 (7-year data).
 We show that parametric fits based on the $\Lambda$CDM model are remarkably close to our nonparametric fits
 in $l$-regions where data are sufficiently precise.
 In contrast,
 the power spectrum for an H$\Lambda$CDM model gets progressively pushed away from our nonparametric fit
 as data quality improves with successive data realizations,
 suggesting incompatibility of this particular cosmological model with respect to the WMAP data sets.
 We present uncertainties on peak/dip locations and heights at the 95\% ($2 \sigma$) level of confidence,
 and show how these uncertainties translate into hyperbolic ``bands"
 on the acoustic scale ($l_A$) and peak shift ($\phi_m$) parameters.
 Based on the confidence set for the 7-year data,
 we argue that the low-$l$ up-turn in the CMB  power spectrum
 cannot be ruled out at any confidence level in excess of about 10\% ($\approx 0.12 \sigma$).
 Additional outcomes of this work are
 a numerical formulation for minimization of a noise-weighted risk function subject to monotonicity constraints,
 a prescription for obtaining nonparametric fits that are closer to cosmological expectations on smoothness,
 and a method for sampling cosmologically meaningful power spectrum variations from the confidence set of a nonparametric fit.
\end{abstract}

\keywords{cosmic background radiation --- cosmological parameters --- Methods: data analysis --- Methods: statistical}

\section{Introduction}
 \label{intro}

 The angular power spectrum of cosmic microwave background (CMB) temperature fluctuations
 is a measurable physical quantity that depends sensitively on the physics of the early universe.
 In particular, the shape of the angular power spectrum and
 the locations and heights of its peaks relate directly to parameters in the underlying cosmological models.
 As such, it has been used extensively as an acid test of the relative merit of competing cosmological models,
 and as a rich source of information about cosmological parameters themselves.

 Traditionally, and almost exclusively, cosmologists have resorted to model-based
 parametric statistical methods for estimating the CMB angular power spectrum from data.
 Parametric regression methods require the functional form of the unknown regression function $f$ to be pre-specified.
 The adjustable parameters in $f$, finite in number that is independent of the data size, are usually estimated by maximizing an appropriate likelihood function or a posterior distribution.
 In the cosmological context, it is conventional to employ
 parametric models that attempt to capture the essential physics of the problem
 via the pre-specified functional form, and any pre-existing knowledge about parameters is
 incorporated in the estimation process via appropriate prior distributions.

 Nonparametric function estimation methods, on the other hand, assume no specific functional form for $f$,
 except for mild regularity conditions such as smoothness assumptions, membership to a function space, etc.
 In this approach,
 the number of parameters that define the unknown regression function $f$ is either infinite or grows proportionally with the data size,
 and the estimate $\widehat{f}_N$ of $f$ is obtained by balancing bias and variance of $\widehat{f}_N$ via optimal smoothing.
 Nonparametric methods are therefore model-independent, and are based on fewer and less restrictive assumptions about $f$.
 This, in turn, implies that any inferences about $f$ made from nonparametric regression analyses
 tend to be more data-driven as opposed to being primarily model-driven.
 In other words, to a greater extent nonparametric function estimation methods
 tend to infer what \emph{is} rather than what \emph{should be}.
 As such, nonparametric regression methods can be meaningfully employed as sanity-enforcing mechanisms
 on parametric analyses, thereby making the conclusions drawn more conservative.
 For example, a feature seen in a parametric fit that survives in a nonparametric analysis
 is likely to be a real and robust feature of the data itself, and not merely an
 artifact that is seen because a parametric model expects it to be there.

 Alternative methodologies, such as the nonparametric methodology \citep{GMN+2004,BSM+2007} used in this work,
 may allow posing inferential questions that are difficult to address using conventional methods.
 For example, this particular nonparametric methodology allows validating model-based,
 parametric fits against the confidence set for the nonparametric fit to the same data,
 possibly to rule them out as candidates for the true but unknown regression function.
 This methodology also has the formal advantage of being able to provide
 realistic uncertainties on \emph{any} number of features of the angular power spectrum
 that are simultaneously valid at the same level of confidence.
 Such desirable features are arguably lacking in most mainstream methodologies,
 including Bayesian, that are commonly used in cosmology.
 (An incisive, insightful, and discerning discussion about the relative merits of this methodology
 over statistical methods conventionally employed in cosmology can be found in the two references
 cited above, and is best read in the original.)

 The four CMB angular power spectrum data sets
 \citep{HSV+2003,HNB+2007,NDH+2009,LDH+2011}
 released by the Wilkinson Microwave Anisotropy Probe (WMAP) \citep{BBH+2003} mission,
 representing cumulative observations at the end of 1, 3, 5, and 7 years of operation,
 present a unique opportunity for statistical analysis.
 For example, till date these are claimed to represent the most precise
 and extensive full-sky CMB measurements ever made
 (only to be superseded by the Planck mission \citep{TMP+2010}).
 The four data sets represent the evolution of data over a period of about a decade,
 thereby making it possible to assess progressive and possibly systematic resolution of features of the spectrum.
 From a statistical perspective, each of these moderately large data sets
 (minimum of 899 data points for the WMAP 1-year release)
 is not only heteroskedastic, but also has substantial correlations that arise in
 typical data pipelines \citep{Tegmark1997,HBB+2003,HWH+2009,JBG+2007,JBD+2011}.

 In this paper, we present a comparative nonparametric analysis of the
 WMAP 1-, 3-, 5-, and 7-year data sets for the angular power spectrum.
 Specifically, for each data realization, we address the following three key questions:
 (a) How well is the angular power spectrum determined by the data alone?
 (b) How well is the $\Lambda$CDM model supported by a model-independent, nonparametric, data-driven analysis?
 (c) What are the realistic uncertainties on peak/dip locations and heights?
 Our analysis is based on a nonparametric function estimation methodology \citep{GMN+2004,BSM+2007},
 which is discussed in Sec.\ \ref{method} together with our extensions;
 i.e.,
 a numerical formulation for minimization of a inverse-noise-weighted risk function subject to monotonicity constraints,
 a prescription for obtaining nonparametric fits that are closer to cosmological expectations on smoothness,
 and a method for sampling cosmologically meaningful power spectrum variations from the confidence set of a nonparametric fit.
 Results are presented in Sec.\ \ref{results}, and conclusion in Sec.\ \ref{conclusion}.

\section{Methodology}
 \label{method}

 The nonparametric function estimation methodology \citep{GMN+2004,BSM+2007} used in this work
 is an extension of the REACT methodology \citep{Beran2000,Beran2000b},
 which is in turn founded on rigorous formal results in \citep{BD1998,Beran1996}.
 Two early papers \citep{WMN+2001,MNG+2002} used this methodology to analyze,
 under the assumption of homoskedasticity,
 a pre-WMAP data set that combined BOOMERanG, MAXIMA, and DASI data sets.
 A generalization of this formalism for dealing with heteroskedasticity
 via an inverse-noise-weighted loss function was developed in \citep{GMN+2004}.
 More recently, using the WMAP 1-year data, \citep{BSM+2007} illustrated how
 confidence intervals on cosmological parameters and boundaries in the cosmological parameter space
 can be inferred from the confidence set for a nonparametric power spectrum fit.

 In this section, we first present an operational outline of this methodology (Sec.\ \ref{method:fit} and \ref{method:cset}).
 This outline is entirely based on \citep{GMN+2004,BSM+2007} and is included here for completeness.
 A pedagogic treatment of the central ideas and a simpler variant of the problem can be found in \citep{Wasserman2006}.
 Specific citations to other sources are provided wherever appropriate.

 Our own numerical formulation of the monotone risk minimization problem,
 where the risk function is derived from an inverse-noise-weighted loss function,
 is presented in Sec.\ \ref{method:minimization}.
 In Sec.\ \ref{method:selection}, we describe a systematic way of obtaining a monotone fit
 with smoothness that meets cosmological expectations.
 In Sec.\ \ref{method:probing_cs}, we describe our method for probing the confidence set;
 this is the basis for the results presented in Sec.\ \ref{results:CoV} and \ref{results:boxes}.

 \subsection{The nonparametric fit}
 \label{method:fit}

 We are given CMB angular power spectrum data of the form
 \begin{equation}
  Y_l = C_l + \epsilon_l
 \end{equation}
 consisting of $N$ data points observed over multipole index range $l_{min} \le l \le l_{max}$.
 Here, $C_l$ stands for the value of the true but unknown power spectrum at multipole index $l$.
 The noise variables $\epsilon_l$ are assumed to have a mean-0 normal distribution
 with \emph{known} covariance matrix $\Sigma$.
 In practice, any reasonable estimate/approximation $\widehat{\Sigma}$ of this covariance matrix,
 such as an inverse Fisher matrix for a pilot parametric fit, can be used in place of $\Sigma$.

 This nonparametric regression method is based on expanding the unknown regression function $f$,
 assumed to belong to an appropriate $L_2$ function space, in a complete orthonormal basis $\{ \phi_j(x) \}$, as
 $$f(x) = \sum_{j=0}^\infty \beta_j \phi_j(x).$$
 A basis that has proven useful in the CMB angular power spectrum context is the cosine basis
 defined over $0 \le x \le 1$:
 \begin{equation}
  \label{eq:cosinebasis}
  \phi_j(x) = \begin{cases} 1 \;\;\;\;\;\;\;\;\;\;\;\;\;\;\;\;\;\; (j=0) \\ \sqrt{2} \cos(j\pi x) \; (j = 1, 2, \ldots) \end{cases}.
 \end{equation}
 Assuming that $f$ is sufficiently smooth, we take $$f(x) \approx \sum_{j=0}^{N-1} \beta_j \phi_j(x),$$
 and estimate it as 
 \begin{equation}
  \label{eq:fhat}
  \widehat{f}_N(x) = \sum_{j=0}^{N-1} \widehat{\beta}_j \phi_j(x).
 \end{equation}
 While the method is asymptotically (i.e., as the data size $N \rightarrow \infty$) basis-independent,
 choice of the basis may matter in any finite-$N$ application;
 see \citep{Beran2000,Beran2000b} for a detailed discussion.
 This basis satisfies a discrete orthogonality property
 when the data $Y_i$ are available over an equispaced grid $\{x_i = ( 2 i + 1 ) / 2 N, 0 \le i \le N-1\}$
 consisting of zeros of $\phi_N(x)$.
 In the CMB context, any contiguous range of $N$ integer multipole indices $l_{min} \le l \le l_{max}$
 can be formally mapped onto this equispaced grid, hence we will not make any categorical distinction
 between data index $i$ and the corresponding multipole index $l$.

 The true angular power spectrum $C_l \equiv f(x_l)$ is estimated
 as $\widehat{C}_l \equiv \widehat{f}_N(x_l)$ via coefficient estimates $\widehat{\beta}_j$,
 which are estimated as
 \begin{equation}
  \label{eq:betahat}
  \widehat{\beta}_j = \lambda_j Z_j,
 \end{equation}
 where
 \begin{equation}
  \label{eq:Z}
  Z_j = {1 \over N} \sum_{i=0}^{N-1} Y_i \phi_j(x_i) = {( U^T Y )_j \over \sqrt{N}}.
 \end{equation}
 Here, $U$ is the orthonormal matrix with elements $U_{ij} = \phi_j(x_i) / \sqrt{N}$
 and $Y \equiv (Y_0,\ldots,Y_{N-1})^T$.
 The task of obtaining coefficient estimates $\widehat{\beta} \equiv ( \widehat{\beta}_0, \ldots, \widehat{\beta}_{N-1} )^T$,
 and thereby the fit $\widehat{f}_N(x_i)$,
 is now relegated to determining the \emph{shrinkage parameters} $\lambda_j$.
 Assuming smoothness for $f$ (which implies a rapid decay of the true coefficients $\beta_j$ with $j$),
 the shrinkage parameters $\lambda_j$ are constrained to be monotonically decreasing
 \begin{equation}
  \label{eq:monotone}
  1 \ge \lambda_0 \ge \lambda_1 \ge \ldots \ge \lambda_{N-1} \ge 0 \;\; \mbox{(Monotone shrinkage)}.
 \end{equation}
 A special, discrete subset of the monotone shrinkage defined above is the
 \emph{nested subset selection} (NSS) shrinkage, defined as
 \begin{equation}
  \label{eq:nss}
  \lambda_j =
  \begin{cases}
   1 \mbox{ for $0 \le j < J$} \\ 0 \mbox{ for $J \le j < N$}
  \end{cases} \mbox{(NSS shrinkage)}.
 \end{equation}
 Either shrinkage type results in selective damping of high-frequency harmonics in the data $Y_i$,
 which results in smoothing of the fit $\widehat{f}_N$.
 A useful interpretation of shrinkage parameter $\lambda_j$ is that
 it represents the \emph{effective degree of freedom} for the $j$th coefficient estimate $\widehat{\beta}_j$.
 One can thus define the effective degrees of freedom (EDoF) for the entire fit $\widehat{f}_N$ as
 \begin{equation}
  \label{eq:edof}
  \mbox{EDoF}(\lambda) = \sum_{j=0}^{N-1} \lambda_j.
 \end{equation}
 This definition follows from the fact that for a linear smoother, EDoF of a fit is formally defined as $\mbox{tr}( \mathcal{H} )$, where $\mathcal{H}$ is the matrix that connects the fitted values $\widehat{Y}$ to the data $Y$ as $\widehat{Y} = \mathcal{H} Y$. For the present nonparametric regression method, $\mathcal{H}=U D U^T$, where $U$ is the orthonormal basis matrix (Eq.\ \ref{eq:Z}), and $D \equiv \mbox{diag}(\lambda_0, \ldots, \lambda_{N-1})$. This implies $\mbox{tr}(\mathcal{H}) = \mbox{tr}(D) = \sum_{i=0}^{N-1} \lambda_i$.

 In the present formalism, the discrepancy between the (unknown) regression function $f$ and its estimator $\widehat{f}_N$
 is measured by the inverse-noise-weighted squared \emph{loss function} $L(\lambda)$, defined as
 $$
  L(\lambda) = \int_0^1 { \left( f(x) - \widehat{f}_N(x) \over \sigma(x) \right)^2 } dx.
 $$
 Here, $\sigma^2(x)$ is the (known) variance of the data $Y$ at $x$,
 which accounts for the heteroskedasticity of the data $Y$.
 The loss $L$ is considered a function of the vector of shrinkage parameters
 $\lambda \equiv (\lambda_0, \ldots, \lambda_{N-1})^T$
 that determine the regression estimator $\widehat{f}_N$ via Eq.\ \ref{eq:betahat}.
 Risk $R(\lambda)$, which is the expected value of $L(\lambda)$,
 can be written as a sum of two non-negative terms; namely,
 $$
  R(\lambda) =   \int_0^1 \left( { f(x) - \mathbb{E}\left( \widehat{f}_N(x) \right) \over \sigma(x) } \right)^2 dx
               + \int_0^1 \mathbb{E}\left[ \left( { \widehat{f}_N(x) - \mathbb{E}\left( \widehat{f}_N(x) \right) \over \sigma(x) } \right)^2 \right] dx.
 $$
 These two terms represent, respectively, the integrated squared bias
 and the integrated variance of $\widehat{f}_N(x)$, both weighed by $1/\sigma^2(x)$.
 Optimal smoothing is achieved, in principle, by minimizing $R(\lambda)$ with respect to $\lambda$.
 Generally speaking, too little smoothing leads to a fit $\widehat{f}_N$ with low bias and high variance,
 and too much smoothing yields a fit with high bias and low variance.
 Minimal risk or optimal smoothing therefore can be thought of
 as a balance between the bias of $\widehat{f}_N$ and its variance.

 The risk function $R(\lambda)$, unfortunately,
 depends on the unknown regression function $f$, and therefore needs to be estimated.
 A particular estimator of this risk,
 which is of the SURE (Stein's unbiased risk estimator \citep{Stein1981}) kind,
 takes the following form:
 \begin{equation}
  \label{eq:riskhat}
  \widehat{R}(\lambda) = Z^T \bar{D} W \bar{D} Z + \mbox{tr}( D W D B ) - \mbox{tr}( \bar{D} W \bar{D} B ),
 \end{equation}
 where
 $Z \equiv (Z_0,\ldots,Z_{N-1})^T$,
 $D \equiv \mbox{diag}(\lambda_0, \ldots, \lambda_{N-1})$, $\bar{D} = I_N - D$,
 $B = U^T \Sigma U / N$ is the covariance of $Z$, and $I_N$ is the $N \times N$ identity matrix.
 The positive (semi)definite weight matrix $W$ is defined as
 \begin{equation}
  \label{eq:W}
  W_{jk} = \sum_l \Delta_{jkl} w_l,
 \end{equation}
 where $w_l$ is the $l$th coefficient in the expansion $(1 / \sigma^2(x)) \approx \sum_{j=0}^{N-1} w_j \phi_j(x)$
 and, for the cosine basis (Eq.\ \ref{eq:cosinebasis}),
 \begin{displaymath}
  \Delta_{jkl} = \int_{0}^{1} \phi_j(x) \phi_k(x) \phi_l(x) \hspace{1 mm} dx
               = \left\{
                  \begin{array}{ll}
                   1,                                                                        & \mbox{if }\#\{j,k,l=0\}=3, \\
                   0,                                                                        & \mbox{if }\#\{j,k,l=0\}=2, \\
                   \delta_{jk}\delta_{0l} + \delta_{jl}\delta_{0k} + \delta_{kl}\delta_{0j}, & \mbox{if }\#\{j,k,l=0\}=1, \\
                   \frac{1}{\sqrt{2}} (\delta_{l,j+k}+\delta_{l,|j-k|})                      & \mbox{if }\#\{j,k,l=0\}=0. \\
                  \end{array}
                 \right.
 \end{displaymath}

 We denote the optimal shrinkage obtained by minimizing risk $\widehat{R}(\lambda)$ by $\widehat{\lambda}$.
 The best NSS shrinkage $\widehat{\lambda}_{NSS}$ is obtained simply by evaluating risk $\widehat{R}(\lambda)$
 for each of the $(N+1)$ NSS shrinkage vectors and choosing the one with the least risk.
 Monotone shrinkage usually results in a lower risk than the NSS shrinkage
 because of the greater freedom available in the monotone set of shrinkages.
 We will discuss risk minimization subject to monotonicity constraints (Eq.\ \ref{eq:monotone}) in Sec.\ \ref{method:minimization}.

\begin{figure}
  \centerline{
  \includegraphics[width=\textwidth]{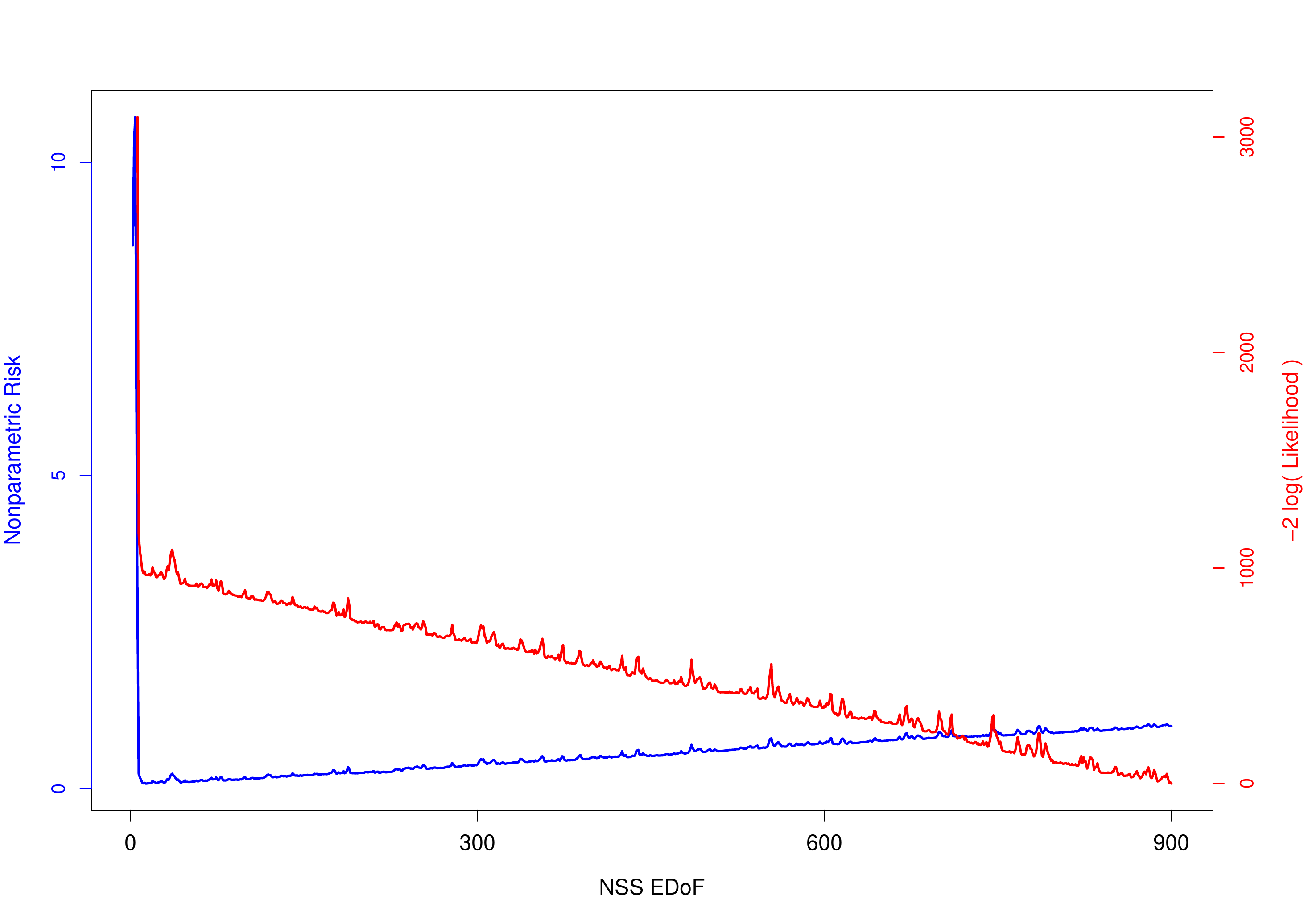}
  }
  \caption{\label{fig:risk_vs_likelihood} Nonparametric risk (red curve) and $-2\log(\mbox{likelihood})$ (blue curve) as functions of the EDoF (Eq.\ \ref{eq:edof}) for the NSS fits to the WMAP 1-year data. This illustrates the contrasting behavior of the two quantities. Optimal smoothing occurs at $\mbox{EDoF} = 12$ where the nonparametric risk attains its minimum over the NSS set of fits. Likelihood function, on the other hand, keeps on improving with the EDoF indefinitely. $\log( \mbox{likelihood} )$ values were computed using the WMAP 1-year likelihood code \citep{VPS+2003}. The blue (left) and red (right) vertical scales on the plot are associated with the nonparametric risk and the $-2\log(\mbox{likelihood})$ respectively. }

 \end{figure}

 Fig.\ \ref{fig:risk_vs_likelihood} illustrates the contrasting behavior
 of the nonparametric risk (red curve) and the WMAP 1-year likelihood function (green curve) \citep{VPS+2003}
 for the WMAP 1-year data \citep{HSV+2003}, as a function of the EDoF of all NSS fits.
 This figure is motivated by the fact that cosmologists, by and large, are better-acquainted with parametric likelihood-based methods.
 Each integer value on the horizontal axis represents one NSS fit,
 from the zero function at $\mbox{EDoF} = 0$ to the fit that simply interpolates through the data ($\mbox{EDoF} = N$).
 Optimal smoothing for NSS shrinkage occurs at $\mbox{EDoF} = 12$ where the nonparametric risk function attains its minimum over the NSS set of fits.
 Likelihood function, on the other hand, keeps on improving with the EDoF indefinitely.

 \subsection{Confidence set around the fit}
 \label{method:cset}

 Conventional regression methods provide a \emph{confidence band} around the fit that quantifies the uncertainty in the fit.
 In contrast, this nonparametric methodology quantifies the uncertainty surrounding the nonparametric fit in the form of an elegant construct,
 namely, a $(1 - \alpha)$ \emph{confidence set} at a pre-specified confidence level $0 \le (1 - \alpha) \le 1$.
 The $(1-\alpha)$ confidence set for the coefficient vector $\beta$ is defined as
 \begin{equation}
  \label{eq:csetD}
  \mathcal{D}_{N,\alpha} = \left\{ \beta: (\beta-\widehat{\beta})^T W (\beta-\widehat{\beta}) \le {r}^2_\alpha \right\},
 \end{equation}
 which is centered at the vector of estimated coefficients $\widehat{\beta}$,
 and the \emph{confidence radius} ${r}_\alpha$ is given by
 \begin{equation}
  \label{eq:cradius}
  {r}^2_\alpha = {\widehat{\tau} z_\alpha \over \sqrt{N}} + \widehat{R}(\widehat{\lambda}).
 \end{equation}
 Here, $z_\alpha$ is the upper $\alpha$ quantile of the standard normal distribution, and
 \begin{equation}
  \label{eq:csettau}
  \widehat{\tau}^2 / N = 2 \mbox{tr}(ABAB) + Z^T Q Z - \mbox{tr}(QB),
 \end{equation}
 with $Q = 4 ( ABA + WDBDW - 2 ABDW )$ and $A = DW + WD - W$.
 In practice, the risk estimator (Eq.\ \ref{eq:riskhat}) and/or $\widehat{\tau}^2$ (Eq.\ \ref{eq:csettau})
 may turn out to be negative for particular data/covariance matrix realizations.
 In such cases, the squared confidence radius (Eq.\ \ref{eq:cradius}) may be negative (or may not be 0 for $\alpha=0$).
 For minimization purposes, the risk estimator (Eq.\ \ref{eq:riskhat}) is adequate and appropriate \citep{BD1998}.
 For confidence radius purposes, we suggest the following modifications to avoid the negativity problem:
 \begin{eqnarray}
  \widehat{R}_+ & = & Z^T \bar{D} W \bar{D} Z + \max\left\{ 0, \mbox{tr}( D W D B ) - \mbox{tr}( \bar{D} W \bar{D} B ) \right\} \nonumber \\
  \widehat{\tau}_+^2 / N & = & 2 \mbox{tr}(ABAB) + \max\left\{ 0, Z^T Q Z - \mbox{tr}(QB) \right\} \\
  {r}^2_{\alpha+} & = & \max\left\{ 0, {\widehat{\tau}_+ z_\alpha \over \sqrt{N}} + \widehat{R}_+ \right\}. \nonumber
 \end{eqnarray}
 At worst, this adjustment will make the confidence radius bigger, resulting into, e.g., wider confidence intervals, but more conservative inferences.
 Similar modifications have been suggested in \citep{BD1998,Beran2000,GMN+2004}.

 The corresponding confidence set on the true regression function $f$ is given by
 \begin{equation}
  \label{eq:csetB}
  \mathcal{B}_{N,\alpha} = \left\{ f(x) = \sum_{j=0}^{N-1} \beta_j \phi_j(x) : \beta \in \mathcal{D}_{N,\alpha} \right\}.
 \end{equation}
 The quadratic form of the inverse noise-weighted loss function
 and the fact that the weight matrix $W$ is positive (semi)definite
 implies that both confidence sets $\mathcal{D}_{N,\alpha}$ and $\mathcal{B}_{N,\alpha}$ are ellipsoidal in shape.
 For any functional $T$ of the spectrum $f$, such as location or height of a peak or a dip,
 the $(1-\alpha)$ confidence interval is defined as
 \begin{equation}
  \label{eq:csetT}
  \mathcal{I}_{N,\alpha} = \left( \min_{f \in \mathcal{B}_{N,\alpha}} T(f), \max_{f \in \mathcal{B}_{N,\alpha}} T(f) \right).
 \end{equation}
 Moreover, prior information that the true regression function $f$ belongs to
 a subset $\mathcal{P}_{N,\alpha}$ of the confidence set $\mathcal{D}_{N,\alpha}$
 (e.g., $f$ has $k$ peaks over the range of $x$-values represented in the data)
 can be incorporated in the analysis by replacing $\mathcal{D}_{N,\alpha}$ with $\mathcal{P}_{N,\alpha} \cap \mathcal{D}_{N,\alpha}$.
 This methodology further provides the formal assurance that, asymptotically,
 \begin{enumerate}
  \item
   $\mathcal{B}_{N,\alpha}$ ($\mathcal{D}_{N,\alpha}$) will contain the true spectrum $f$ (true coefficient vector $\beta$) with probability $\ge (1-\alpha)$, and
  \item
   confidence intervals $\mathcal{I}_{N,\alpha}$ on \emph{any number of} functionals $T(f)$,
   computed from the confidence set $\mathcal{B}_{N,\alpha}$,
   will be \emph{simultaneously valid} at the same confidence level $(1-\alpha)$,
   and that these will trap their corresponding true but unknown values with probability $\ge (1-\alpha)$.
 \end{enumerate}

 \subsection{Risk minimization subject to monotonicity constraints}
 \label{method:minimization}

 In this section, we show how the risk function $\widehat{R}(\lambda)$ (Eq.\ \ref{eq:riskhat}) can be minimized subject to
 the monotonicity constraints $1 \ge \lambda_0 \ge \lambda_1 \ge \ldots \ge \lambda_{N-1} \ge 0$.
 The risk function corresponding to the unweighted loss function ($W = I_N$) has a simple weighted-sum-of-squares form,
 and can be minimized exactly and efficiently using the pooled adjacent violators (PAV) algorithm \citep{RWD1988}.
 While the risk function corresponding to the inverse-noise-weighted loss function ($W \ne I_N$) is still quadratic in $\lambda$,
 it can no longer be expressed as a weighted sum-of-squares, and the PAV algorithm cannot be used to minimize it.

\begin{figure}
  \centerline{
  \includegraphics[width=0.5\textwidth]{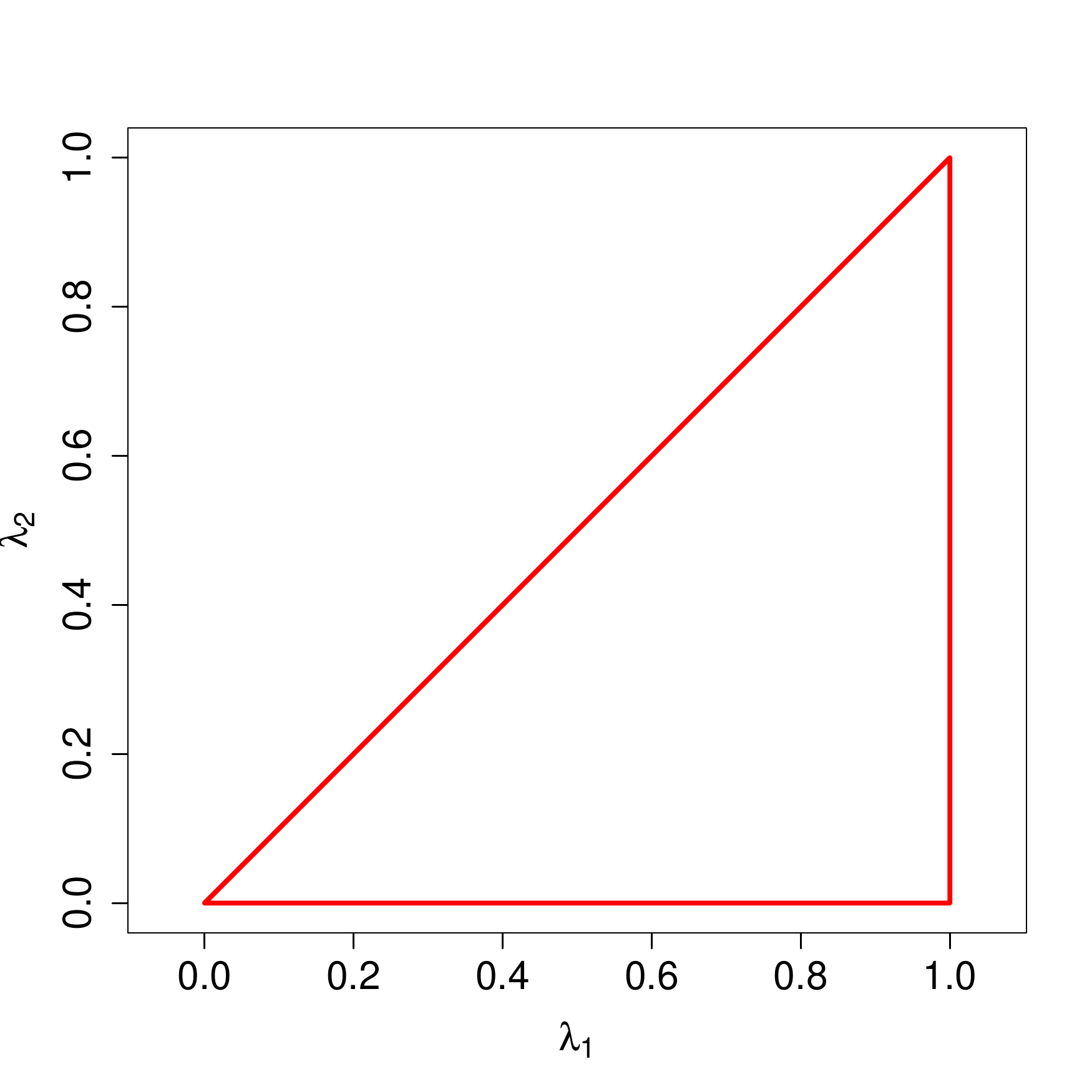}
  \includegraphics[width=0.5\textwidth]{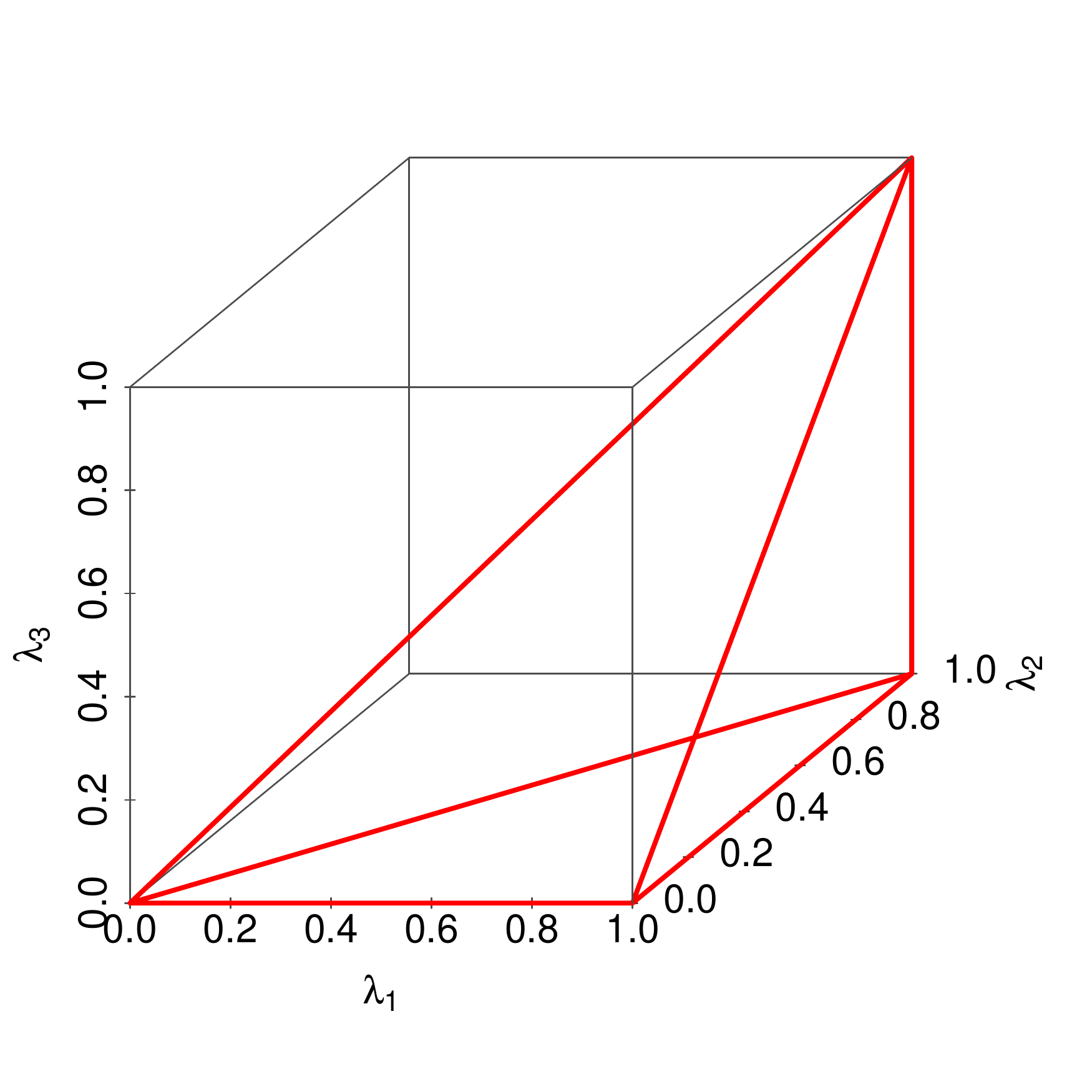}
  }
  \caption{\label{fig:monotone} Trianguloid-shaped admissible regions, marked by red lines, for the monotonicity constraint $1 \ge \lambda_0 \ge \lambda_1 \ge \ldots \ge \lambda_{N-1} \ge 0$ for $N = 2$ (left) and 3 (right). The $(N+1)$ vertices of the trianguloid correspond to the $(N+1)$ NSS fits, with the origin corresponding to the zero function, and the vertex $(1,1,\ldots,1)$ corresponding to the function that exactly interpolates through the data. Surfaces with constant value $p$ of EDoF (Eq.\ \ref{eq:edof}) are hyperplanes of the form $\sum_{i=0}^{N-1} \lambda_i = p$.}
 \end{figure}

 It can be shown that, disregarding terms that do not depend on $\lambda$,
 the risk function $\widehat{R}(\lambda)$
 (Eq.\ \ref{eq:riskhat}) can be written as
 $$
  \widehat{R}(\lambda) = {1 \over 2} \lambda^T H \lambda - \lambda^T h,
 $$
 where
 $H_{jk} = 2 z_j z_k W_{jk}, h = ( H - V ) (1,1,\ldots,1)^T,$ and $V_{jk} = 2 W_{jk} B_{kj}$.
 $H$ and $V$ are both manifestly symmetric.
 Positive (semi)definiteness of $W$ implies that $H$ is a positive (semi)definite matrix,
 implying that $\widehat{R}(\lambda)$ is a convex function.
 The system of linear inequality constraints $1 \ge \lambda_0 \ge \lambda_1 \ge \ldots \ge \lambda_{N-1} \ge 0$
 implies that the constrained region (Fig.\ \ref{fig:monotone}) has a convex trianguloidal shape determined by flat surfaces.
 The original risk minimization problem can therefore be formulated as the following equivalent convex quadratic minimization problem:
 \begin{eqnarray}
  \label{eq:riskmin}
  \mbox{Minimize}   & & \widehat{R}(\lambda) = {1 \over 2} \lambda^T H \lambda - \lambda^T h \nonumber \\
  \mbox{subject to} & & C \lambda\le (0,0,\ldots,0)^T \\
  \mbox{and}        & & 0 \le \lambda_i \le 1 \;\; \mbox{for all $i$,} \nonumber
 \end{eqnarray}
 where $C$ is the $(N-1) \times N$ matrix
 $$
  C = \left[
       \begin{array}{rrrrrrr}
             -1 &      1 &      0 & \ldots &      0 &      0 &      0 \\
              0 &     -1 &      1 & \ldots &      0 &      0 &      0 \\
         \vdots & \vdots & \vdots & \vdots & \vdots & \vdots & \vdots \\
              0 &      0 &      0 & \ldots &     -1 &      1 &      0 \\
              0 &      0 &      0 & \ldots &      0 &     -1 &      1 \\
       \end{array}
  \right].
 $$
 An additional linear inequality or equality constraint of the form
 \begin{equation}
  \label{e:edof_q}
  \sum_{i=0}^{N-1} \lambda_i \le q \;\;\; \mbox{ or } \;\;\;
  \sum_{i=0}^{N-1} \lambda_i  = q,
 \end{equation}
 where $q$ constrains the EDoF of the fit ($0 < q \le N$),
 can easily be accommodated in this formulation.
 This reformulation of the monotone risk minimization problem
 makes it possible to use standard minimization methods \citep{PD1978,Powell1985,GI1983} and
 computational tools \citep{Schittkowski2007,Vanderbei1999}
 for convex quadratic minimization problems subject to linear constraints and simple bounds.

 \subsection{Obtaining a smoother monotone fit that is closer to cosmological expectations}
 \label{method:selection}

\begin{figure}
  \begin{center}
   \includegraphics[height=0.5\textheight]{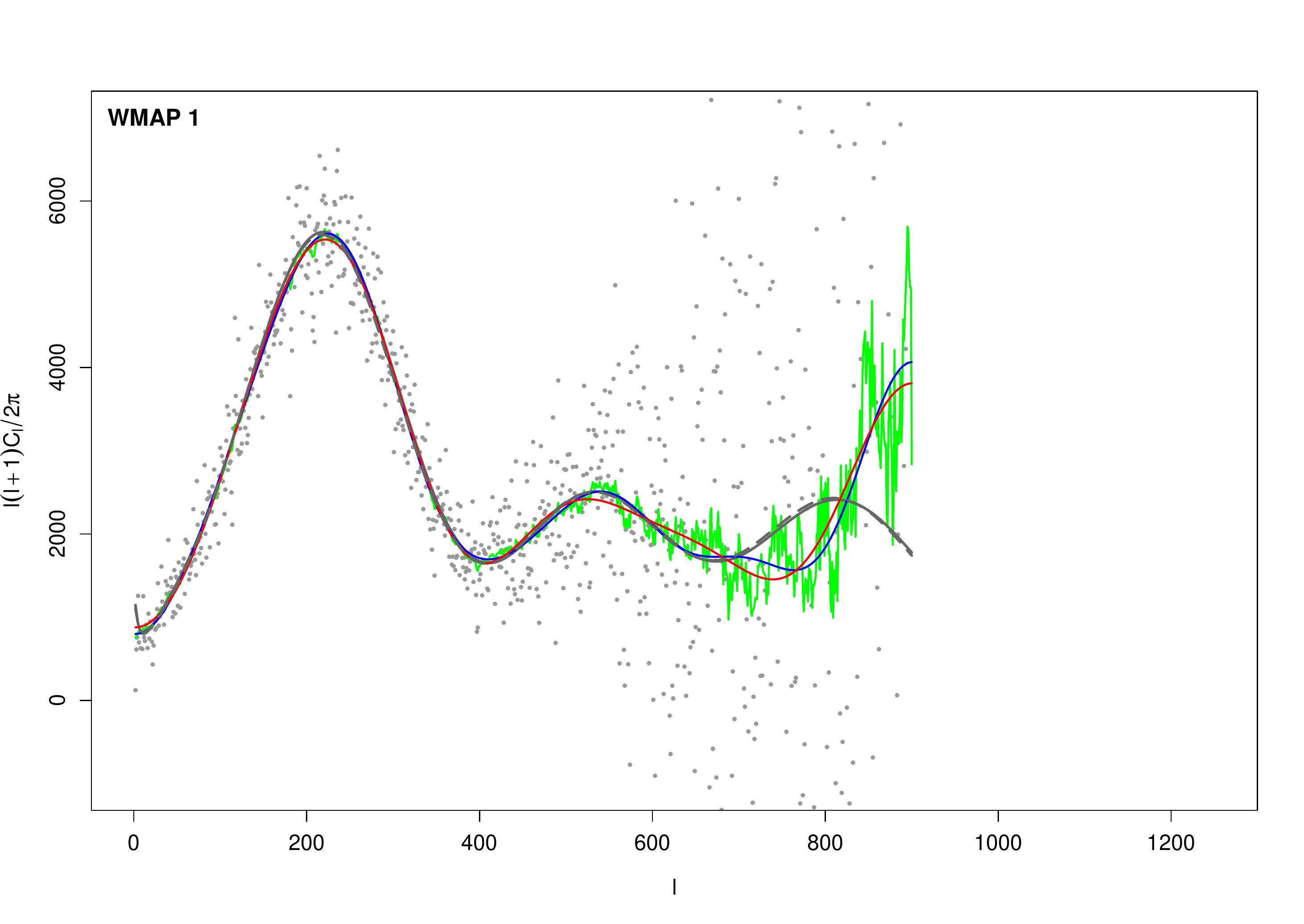}

   \vspace{-4em}
   \includegraphics[height=0.5\textheight]{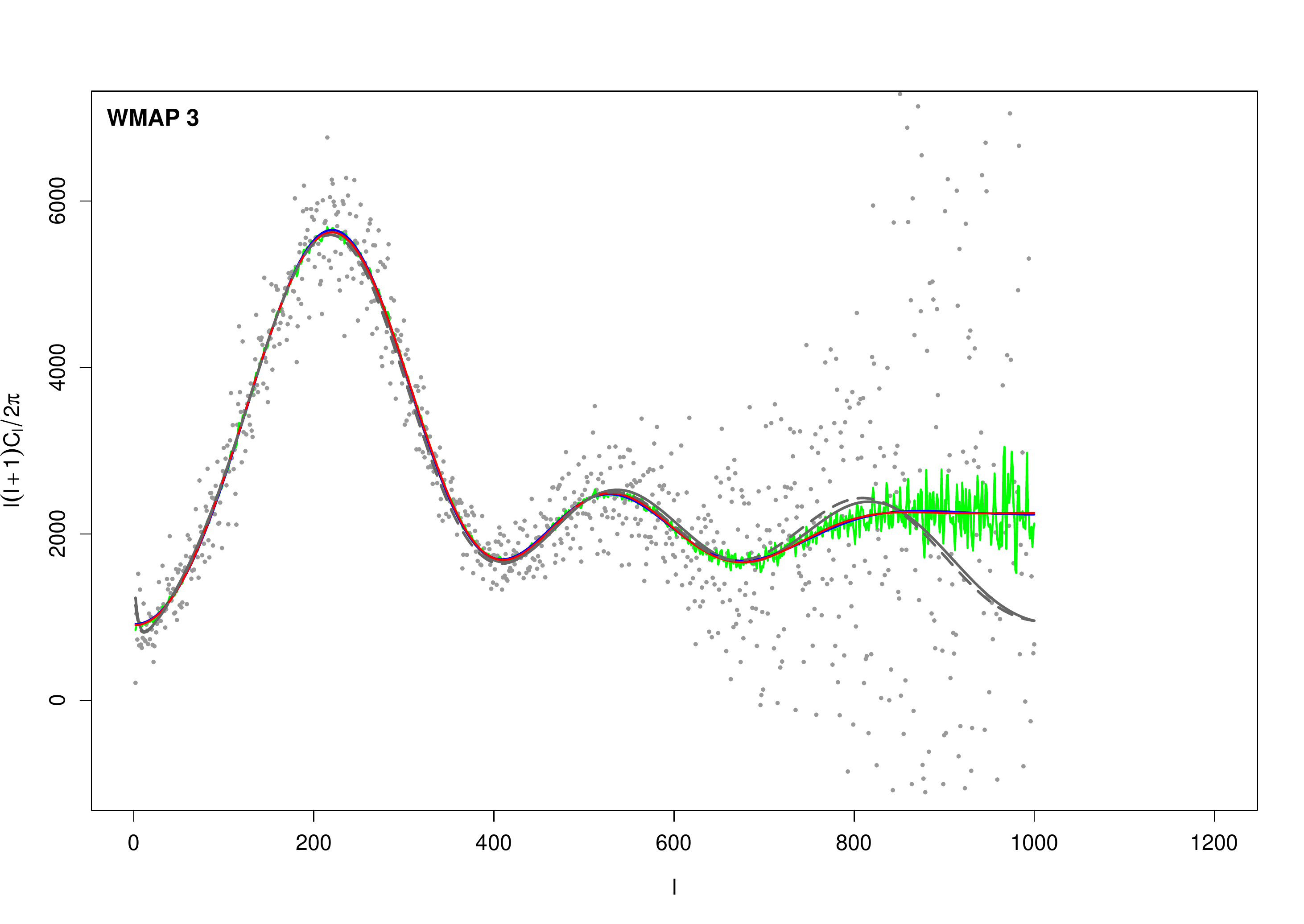}
  \end{center}
  \caption{\label{fig:fits1} Nonparametric fits for WMAP 1- (top) and 3-year (bottom) data sets. $x$- and $y$-ranges are identical across plots. Green: full-freedom monotone fit ($\mbox{EDoF}\approx 80.2,76.5$ respectively); blue: NSS fit ($\mbox{EDoF}=12,10$ respectively); red: restricted-freedom monotone fit ($\mbox{EDoF}\approx 9.4,9.5$ respectively); solid gray: best $\Lambda$CDM-based parametric fit; dashed gray: power spectrum for an H$\Lambda$CDM model. See Table \ref{tab:lambda} for details of model-based power spectra.}
 \end{figure}

\begin{figure}
  \begin{center}
   \includegraphics[height=0.5\textheight]{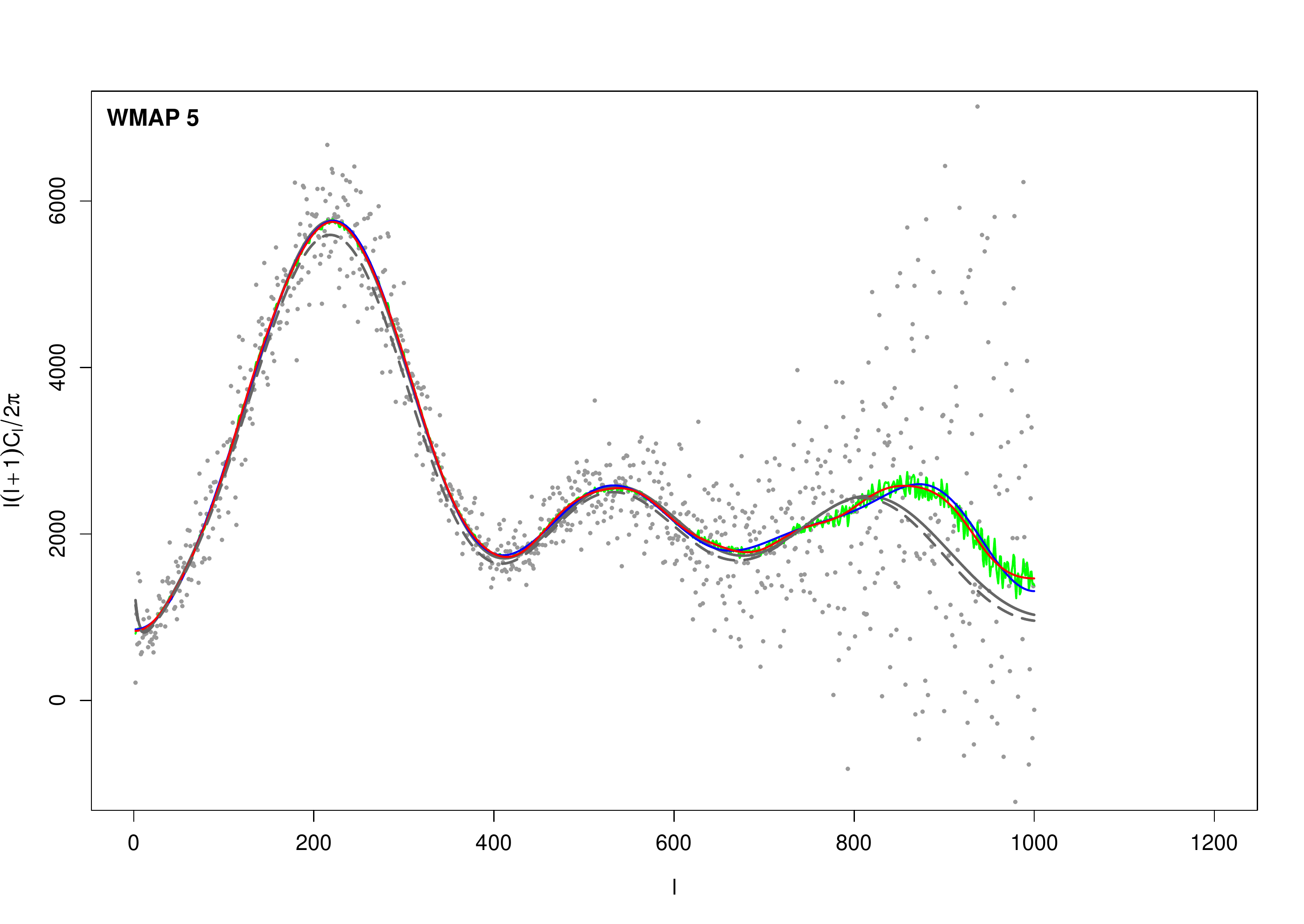}

   \vspace{-4em}
   \includegraphics[height=0.5\textheight]{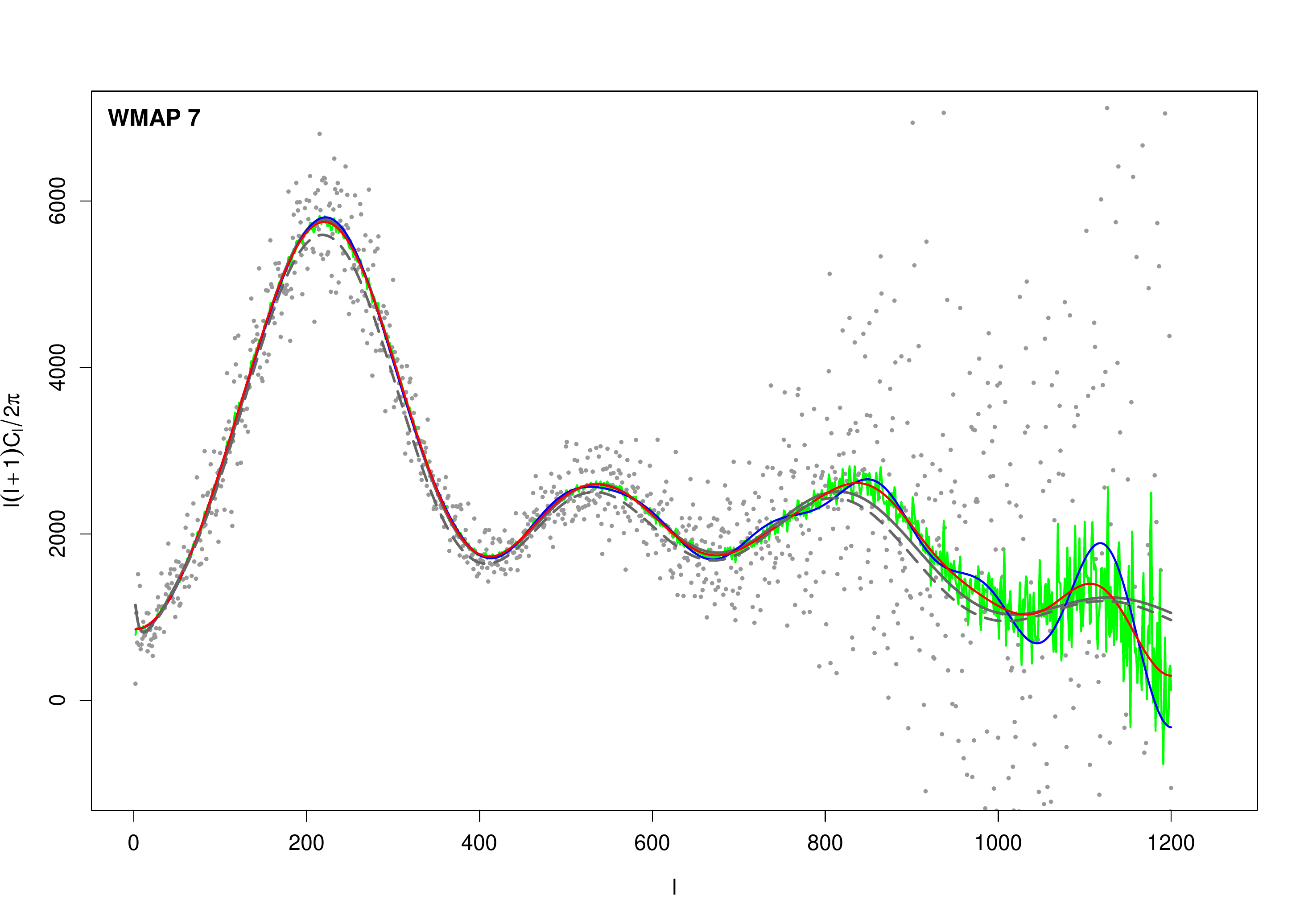}
  \end{center}
  \caption{\label{fig:fits2} Nonparametric fits for WMAP 5- (top) and 7-year (bottom) data sets. $x$- and $y$-ranges are identical across plots. Green: full-freedom monotone fit ($\mbox{EDoF}\approx 60.4,102.9$ respectively); blue: NSS fit ($\mbox{EDoF}=13,20$ respectively); red: restricted-freedom monotone fit ($\mbox{EDoF}\approx 14.4,14.1$ respectively); solid gray: best $\Lambda$CDM-based parametric fit; dashed gray: power spectrum for an H$\Lambda$CDM model. See Table \ref{tab:lambda} for details of model-based power spectra.}
 \end{figure}

 To motivate the discussion in this section, consider the full-freedom monotone fit to the WMAP data sets obtained as above
 (green curves in Fig.\ \ref{fig:fits1} and \ref{fig:fits2}).
 By \emph{full-freedom monotone fit}, we mean the fit that minimizes risk over the
 entire monotone-admissible region (Fig.\ \ref{fig:monotone})
 without any restriction on the EDoF of the fit.
 This fit turns out to be quite wiggly especially at the high-$l$ end because of the high noise variance here.
 Such wiggliness implies presence of high-frequency components in the fit, which in turn implies a large number of EDoF in the fit.

 Without cosmological pre-conditioning (i.e., from a completely agnostic and data-driven perspective)
 and when viewed in relation to the data,
 it is clear that this fit is not at all unreasonable, given the high noise levels at the high-$l$ end.
 However, all cosmological models suggest far smoother shapes for the angular power spectrum.
 In the context of the present methodology, one candidate for a smoother fit is the NSS fit
 (i.e., one that has the minimal risk over the set of $(N+1)$ NSS fits).
 Indeed, this possibility has been exploited, e.g., in \citep{GMN+2004}.
 However, the NSS fit (see the blue curve in Fig.\ \ref{fig:fits1} and \ref{fig:fits2})
 may also turn out to be somewhat unsatisfactory with respect to cosmological expectations
 (and, some times, also with respect to trends reflected in the full-freedom monotone fit).
 This is primarily because of the limited freedom available in the NSS class.
 Notice again that the NSS fit is not entirely unreasonable from an agnostic viewpoint.

 The monotone set, on the other hand, offers the possibility of harnessing local minima
 in the risk function that are constrained to lie in appropriate ``smoother" subsets of the full monotone set.
 This may be achieved in two distinct ways that may be combined for greater effect:
 \begin{enumerate}
  \item
   By imposing one of the additional constraints (Eq.\ \ref{e:edof_q}) on the EDoF of the fit.
   Examples of such restricted-freedom monotone fits are the red curves in Fig.\ \ref{fig:fits1} and \ref{fig:fits2}.
  \item
   By truncating the expansion (Eq.\ \ref{eq:fhat}) to $p$ number of coefficients $(p < N)$
   and then performing monotone risk minimization over this subset of the full monotone set.
 \end{enumerate}
 In practice, such smoother restricted-freedom monotone fit can be obtained by gradually reducing
 the value of $q$ (Eq.\ \ref{e:edof_q}) starting from the EDoF of the NSS fit
 until all low-amplitude, high-frequency wiggles in the fit disappear.
 Generally, the resulting fit has a lower risk than the NSS fit with $\mbox{EDoF} = q$,
 and is manifestly consistent with trends captured by the full-freedom monotone fit.
 We find it useful to present (or consider) all three fits
 (NSS, full-freedom monotone, and restricted-freedom smoother monotone) together:
 This helps build a realistic picture about estimated trends in the data,
 and thereby about the shape of the underlying true spectrum.
 Like the NSS fit, the smoother restricted-freedom monotone fit will generally be more biased than the full-freedom monotone fit.
 This greater bias, however, is partially compensated for by a larger risk
 which results in a larger confidence radius value (Eq.\ \ref{eq:cradius}),
 a larger confidence set, and therefore more conservative inferences.

 \subsection{Probing the confidence set for uncertainties on features of the fit}
 \label{method:probing_cs}

 In this paper, we need to probe the confidence set for a fit for two purposes:
 (a) for validating cosmological models against a nonparametric fit (see Sec.\ \ref{results:model_validation}).
 (b) for finding the uncertainties on specific features of the fit such as peak heights and locations, and
 Below we describe our method to scan and sample the confidence set for the latter.
 Our particular method for probing the confidence set for determining uncertainties on
 features of the fit is based on the following observations:
 \begin{enumerate}
  \item
   The confidence set $\mathcal{D}_{N,\alpha}$ on the vector of coefficients $(\beta_0,\ldots,\beta_{N-1})$,
   by construction, is centered at the vector of estimates $(\widehat{\beta}_0, \ldots, \widehat{\beta}_{N-1})$.
  \item
   The confidence interval defined in Eq.\ \ref{eq:csetT} requires locating extreme variations
   in any functional $T$ of the power spectrum ${f}$; e.g., location or height of a peak.
   The largest possible variations in $T$ will be located as far away from
   the center of the confidence set as possible, i.e., on its surface.
  \item
   Cosmologically meaningful and sufficiently smooth variations in the fitted spectrum $\widehat{C}_l$
   are most likely to be located in the projection of the full confidence set onto the lowest $M$ dimensions.
 \end{enumerate}

 We therefore generate a uniform sample from the projection of the full confidence set
 surface onto the lowest $d$ dimensions, where $2 \le d \le M$, with $M \lesssim 23$.
 For convenience, we use the smoother restricted-freedom monotone fit for this purpose,
 with the justification that the confidence set corresponding to the full-freedom monotone fit
 happens to be nested inside that for this fit, for all four data realizations.
 The Cholesky factorization $W = u^T u$ is used to transform the original confidence ellipsoid $\mathcal{D}_{N,\alpha}$
 (whose principal axes may not be aligned with coordinate directions in the $\beta$-space)
 into a sphere of the form $\{ \psi: \Vert \psi - \widehat{\psi} \Vert^2 \le r^2_\alpha \}$, where
 $\psi = u \beta$ and $\widehat{\psi} = u \widehat{\beta}$.
 The surface of this $\psi$-sphere can be efficiently sampled
 with uniform density using a standard algorithm (Sec.\ 3.4.1.E.6 on p.130 of \citep{Knuth1981}),
 and then transformed back into the $\beta$-space,
 preserving uniformity of density because of the linearity of the transformation.
 From a sufficiently large sample of such variations of the power spectrum,
 we further selected those functions for which successive peaks and dips
 are separated by at least 50 multipole moments $l$.
 This cosmologically-motivated selection criterion ensures that
 (a) the sampled functions are sufficiently smooth, and
 (b) high-frequency wiggles are not counted as peaks or dips when estimating uncertainties on locations and heights of peaks and dips (see Sec.\ \ref{results:boxes}).
 Based on cosmological considerations, we restricted the search
 to functions with 3 peaks (WMAP 1-, 3-, and 5-year data) or 4 peaks (7-year data).
 The set of functions thus sampled is used to estimate uncertainties on specific features of the fit.
 As an aside, we note that the confidence set construct and the formal guarantees related to confidence intervals (Eq. 14) do not necessarily imply a uniform density over the confidence set;
 uniform sampling is used here as a convenient computational device for scanning the confidence set surface in an unbiased fashion.

 \section{Results and Discussion}
 \label{results}

 The four WMAP angular power spectrum data sets used in this work,
 $\Lambda$CDM parametric fits for the CMB angular power spectrum,
 and likelihood codes that produce their respective Fisher (inverse covariance) matrices
 are obtained from the WMAP data archive \texttt{http://lambda.gsfc.nasa.gov/}.
 For all four data realizations, it turned out that
 (a) the weight matrix $W$ (Eq.\ \ref{eq:W}) is numerically positive definite, and
 (b) the confidence set for the full-freedom monotone fit is completely nested inside that for the smoother restricted-freedom monotone fit.
 It is worth noting that our nonparametric fits and confidence sets are not too sensitive to the details of the covariance matrix
 (this was also pointed out in \citep{BSM+2007}).
 Most computations were done using the \texttt{R} statistical computing environment \citep{R}.
 We used the \texttt{QL} codes \citep{Schittkowski2007} for the monotone risk minimization problem (Eq.\ \ref{eq:riskmin}).
 Our nonparametric fits, obtained using the method outlined in Sec.\ \ref{method:minimization},
 are shown in Fig.\ \ref{fig:fits1} and \ref{fig:fits2} for the WMAP 1-, 3-, 5-, and 7-year data sets.

 \subsection{How well is the power spectrum determined by data alone?}
 \label{results:CoV}

 Considering that the noise in all four data sets is highly heteroskedastic and noise levels are especially high for large $l$,
 it would be useful to make an assessment of how such noise in the data affects local uncertainties in the fitted spectrum,
 and to quantify how well the angular power spectrum value at each $l$ is determined by the data.

 To this end, we compute, for each data set, the approximate 95\% confidence interval on each $\widehat{C}_l$
 using 5000 function variations from the confidence set as outlined in Sec.\ \ref{method:probing_cs}.
 The length of this vertical confidence interval at given $l$,
 divided by the absolute value of the fit, $\vert \widehat{C}_l \vert$,
 provides an approximate indication of how well each $\widehat{C}_l$ is determined
 via the following interpretation:
 a value $\ll$ 1 indicates that the fit is well determined by the data,
 whereas values $\gtrsim$ 1 imply that the data contain little or no information about the height of the power spectrum.
 This approach, which is inspired by the boxcar probe approach of \citet{GMN+2004},
 has the practical advantage of not having adjustable parameters (i.e., the boxcar width) in the procedure.

\begin{figure}
  \centerline{
  \includegraphics[width=\textwidth]{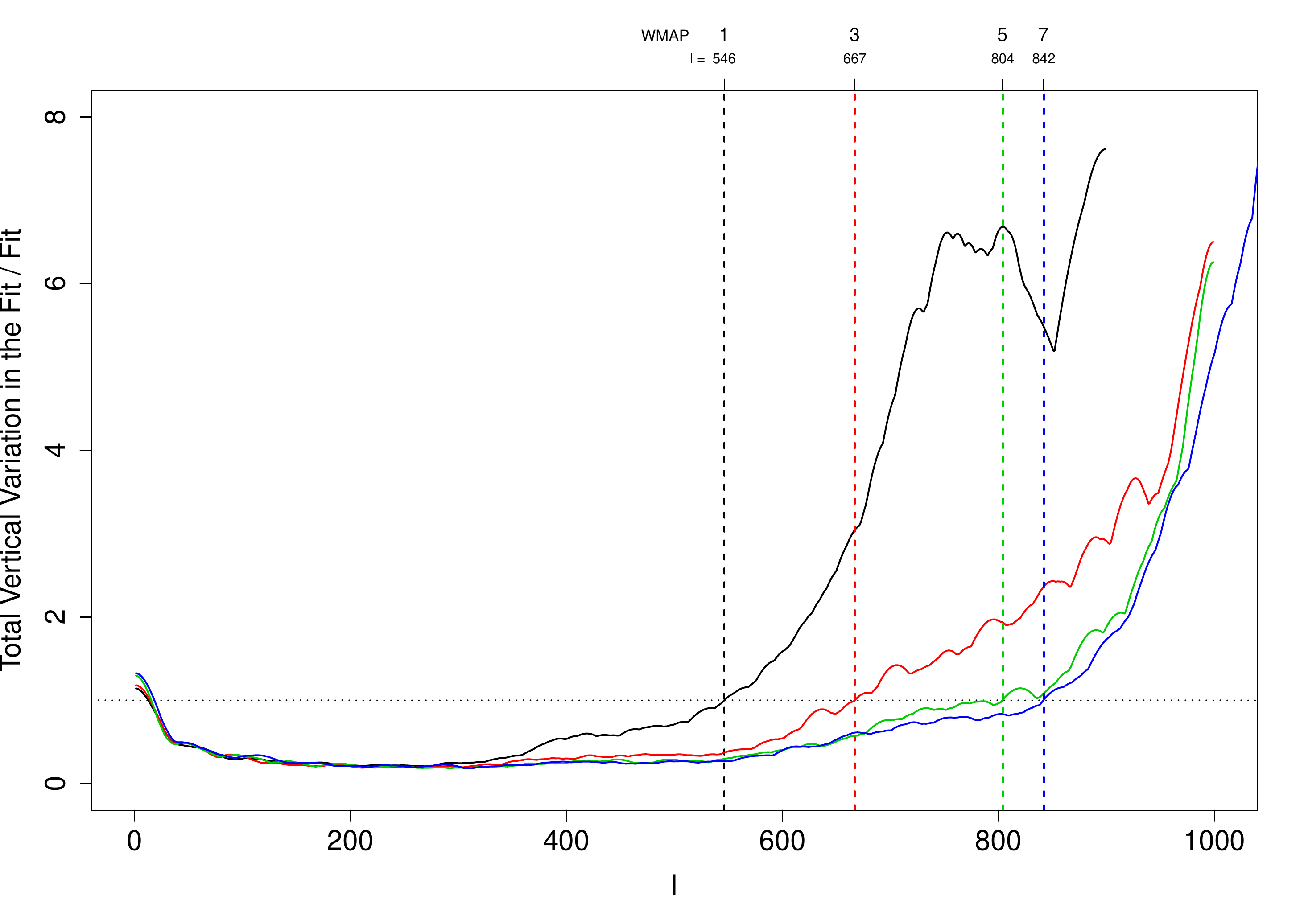}
  }
  \caption{\label{fig:CoV} The results of a probe of the confidence sets for the WMAP 1- (black), 3- (blue), 5- (red), and 7-year (green) nonparametric restricted-freedom monotone fits (Fig.\ \ref{fig:fits1} and \ref{fig:fits2}) to determine how well the angular power spectrum is determined by the data alone. The quantity plotted for each data realization is the total vertical variation at each $l$ within the respective 95\% ($2 \sigma$) confidence set, divided by the absolute value of the fit. This quantity is an approximate measure of how well the angular power spectrum is determined by the data: Values $\ll 1$ indicate that the fit is tightly determined by the data, whereas values $\gtrsim$ 1 indicate that the data contain little or no information about the height of the angular power spectrum for that $l$. Disregarding the low-$l$ region, the color-coded vertical lines indicate the approximate $l$-value at which each curve rises above 1.}
 \end{figure}

 In Fig.\ \ref{fig:CoV}, we plot this height, scaled by the value of the fit, as a function of the multipole index $l$,
 for all four data realizations.
 We see that the range of $l$-values over which the fit is well-determined
 expands consistently between 1-, 3- and 5-year data realizations,
 from $l\approx 546$ (1-year),
 to $l\approx 667$ (3-year),
 to $l\approx 804$ (5-year).
 On the other hand, while the $l$-range of the data expanded substantially between WMAP 5 and 7,
 the information contained in the data does not appear to have grown proportionately beyond $l \approx 842$ for the 7-year data.

 \subsection{How well is the $\Lambda$CDM model supported by a model-independent, nonparametric, data-driven analysis?}
 \label{results:model_validation}

 In each of the four plots in Fig.\ \ref{fig:fits1} and \ref{fig:fits2},
 we have also included parametric fits based on the
 $\Lambda$CDM \citep{HSV+2003,HNB+2007,NDH+2009,LDH+2011} and H$\Lambda$CDM
 \citep{PHK+1995,PG2001}
 models (see figure caption for details).
 The parametric $\Lambda$CDM-based fits turn out to be quite close to the respective nonparametric fits
 wherever the data are precise.
 This is remarkable considering that our nonparametric method does not rely on any cosmologically-motivated
 prior information whatsoever.
 Moreover, the parametric $\Lambda$CDM-based fit appears to get closer to the respective nonparametric fit
 across the four WMAP data releases.
 The closeness of a parametric fit $(C_{l_{min}}, \ldots, C_{l_{max}})$ to
 the corresponding nonparametric fit $(\widehat{C}_{l_{min}}, \ldots, \widehat{C}_{l_{max}})$
 can be measured through the distance
 $$
  d(C,\widehat{C}) \approx \sqrt{ {1 \over N} \sum_{l=l_{min}}^{l_{max}} \left( C_l - \widehat{C}_l \over \sigma_l \right)^2  }.
 $$
 Using Eq.\ \ref{eq:cradius}, this distance can be further expressed as the smallest confidence level $\alpha$
 beyond which the parametric fit is rejected as a candidate for the true spectrum.

 Table \ref{tab:lambda} lists distances of parametric fits based on the two cosmological models
 ($\Lambda$CDM and H$\Lambda$CDM) from the nonparametric full-freedom monotone fit
 for the corresponding data realization (see caption for specific details and description).
 The progression of distance values between a parametric $\Lambda$CDM fit
 and the corresponding nonparametric fit
 clearly shows that the two are getting closer as the data become precise.
 In contrast to this, the angular power spectrum generated by the particular H$\Lambda$CDM model considered,
 which is almost as strong a contender for the true power spectrum as the $\Lambda$CDM fit with respect to the 1-year confidence set, is progressively pushed away to the boundary of the 95\% confidence set for the 7-year confidence set.
 Visually, this trend can be understood on the basis of the differences between the H$\Lambda$CDM power spectrum
 and the nonparametric fit (e.g., differences in the heights of the first peak; see Fig.\ \ref{fig:fits1} and \ref{fig:fits2}) that result into pushing this particular H$\Lambda$CDM model out of the confidence set.
 Given the formal guarantees of this methodology (Sec.\ \ref{method:cset}),
 the WMAP 7-year data thus rules out, at $\approx 95\%$ confidence level, the particular H$\Lambda$CDM model considered.

 \subsection{Uncertainties on locations and heights of peaks and dips}
 \label{results:boxes}

 We now consider the problem of determining uncertainties on the locations and heights
 of peaks and dips in the nonparametrically fitted spectrum.
 The motivation for this exercise comes from the fact that peak locations and heights
 contain valuable information about cosmological models and parameters \citep{DL2002,DNA2003}.

 Following the prescription outlined in Sec.\ \ref{method:probing_cs}, we sampled a set of 5000 function variations
 from the confidence set for each data realization.
 Peak and dip locations and heights were recorded for each peak and dip over this set of functions.
 This results into an empirical scatterplot that is indicative of the joint distribution
 of location and height for each peak or dip, under the assumption of uniform surface density
 on the confidence set $\mathcal{D}_{N,\alpha}$.

\begin{figure}
  \centerline{
  \begin{tabular}{c|c}
   \includegraphics[width=0.5\textwidth]{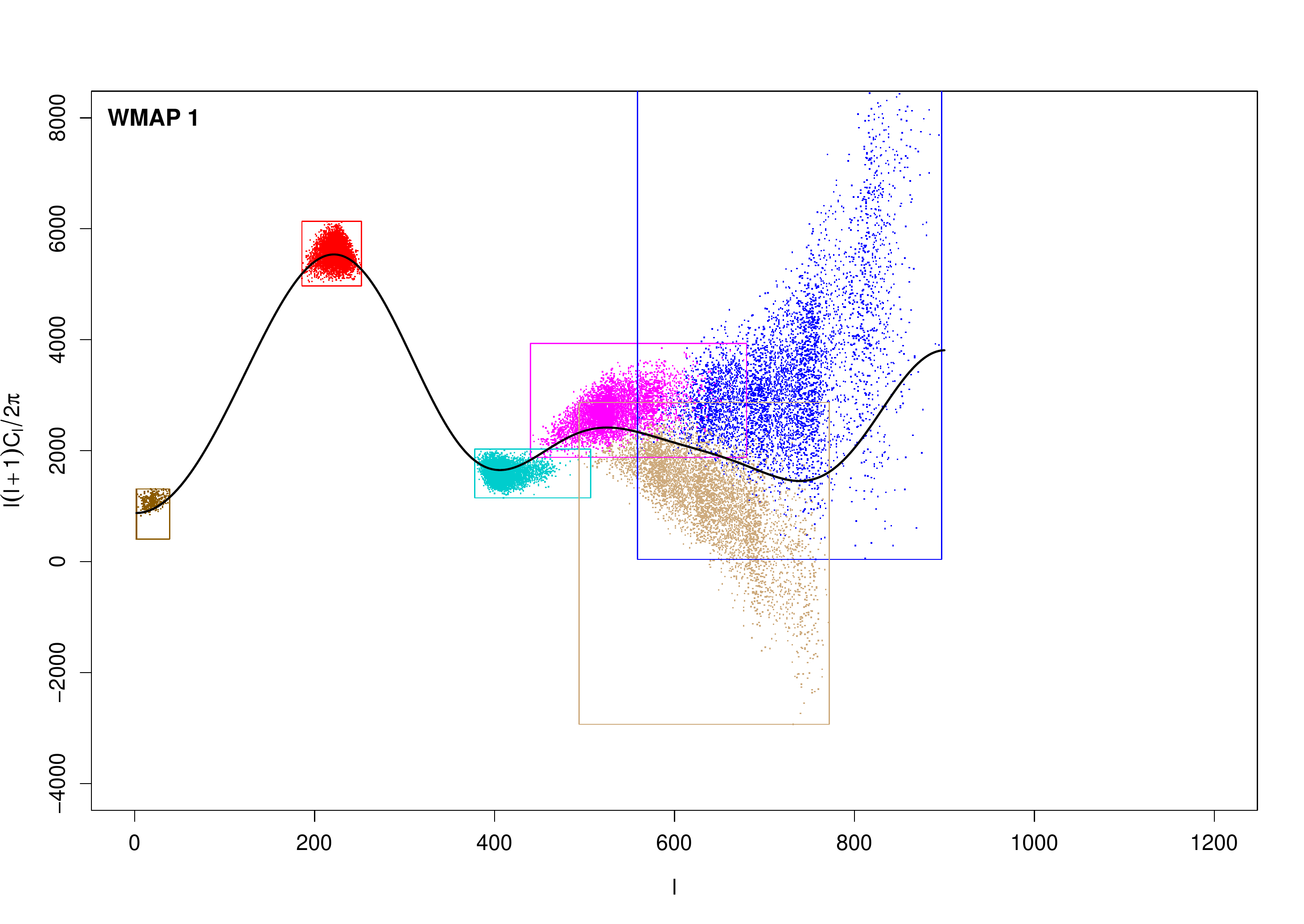}
   &
   \includegraphics[width=0.5\textwidth]{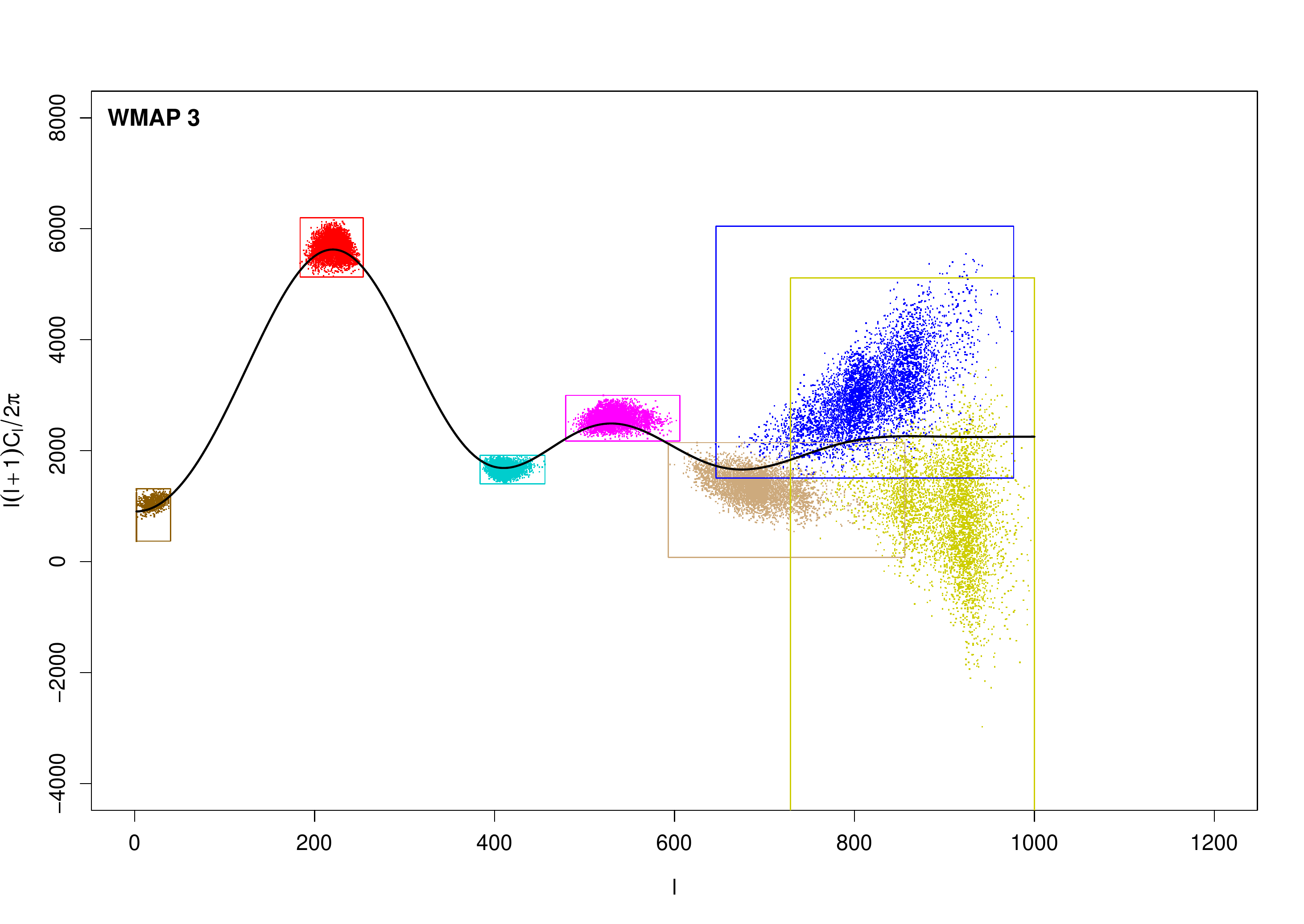}
   \\
   \hline
   \includegraphics[width=0.5\textwidth]{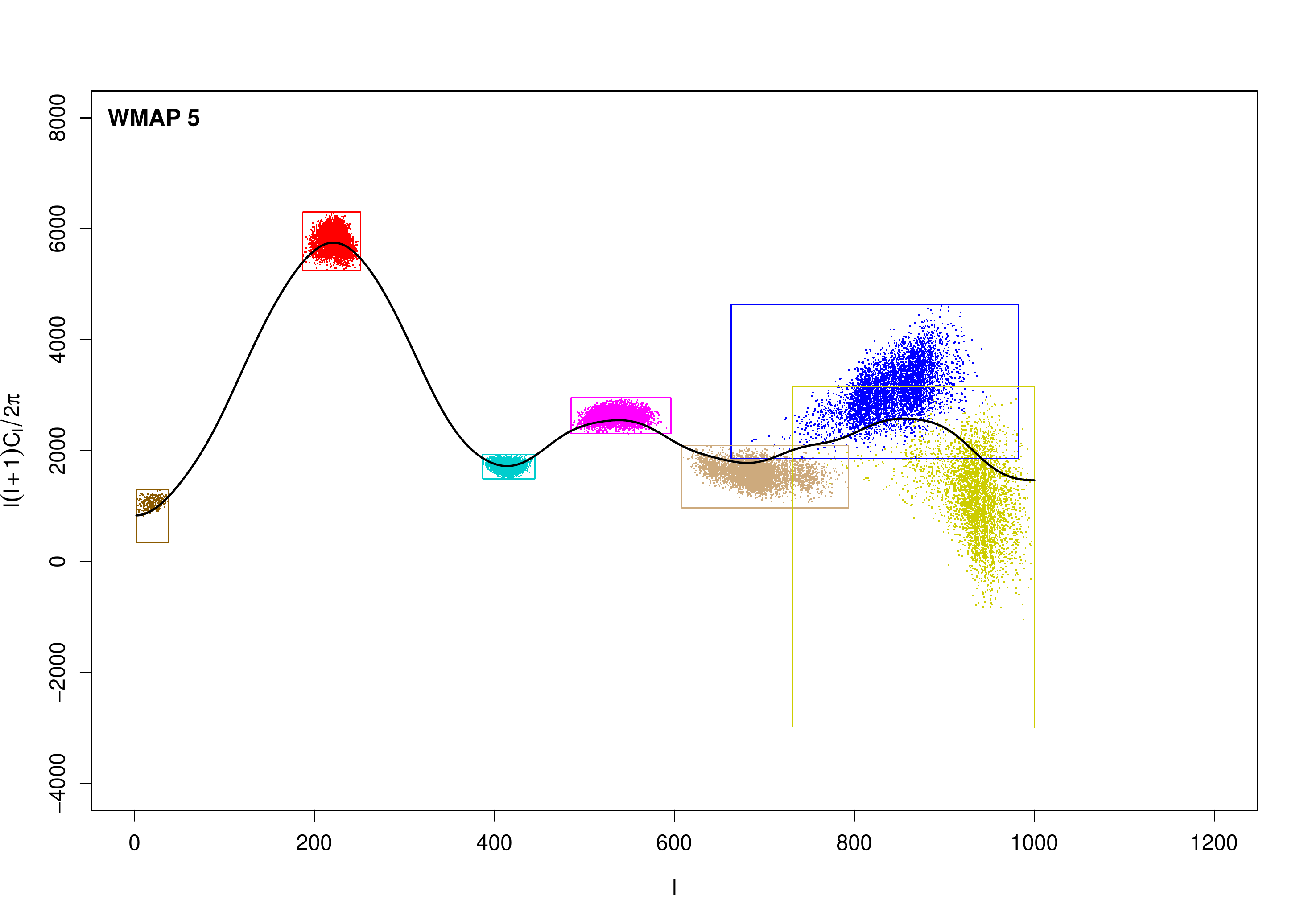}
   &
   \includegraphics[width=0.5\textwidth]{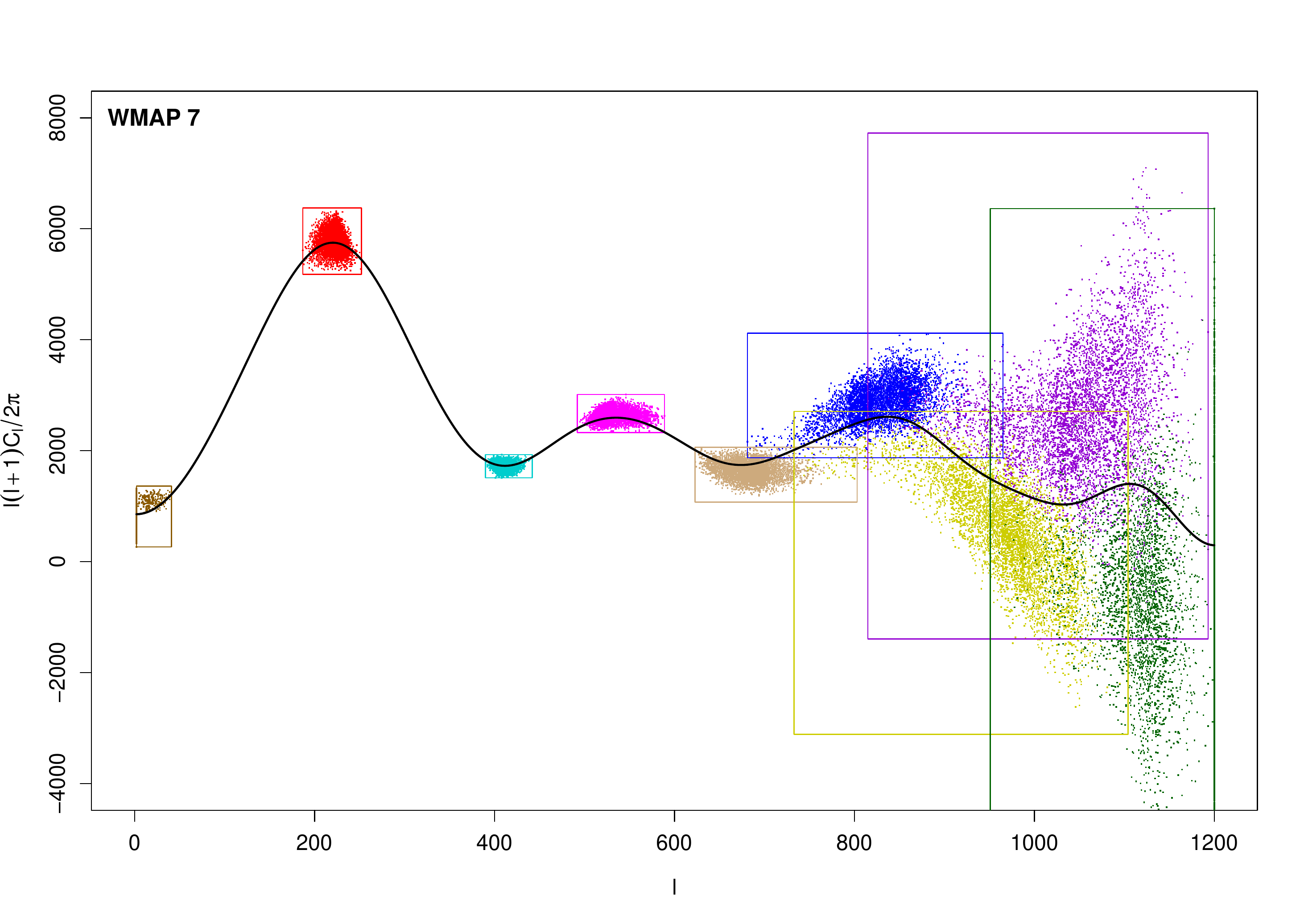}
  \end{tabular}
  }
  \caption{\label{fig:pd} Nonparametric uncertainties on peak and dip locations and heights for the WMAP 1- (top left), 3- (top right), 5- (bottom left), and 7-year (bottom right) data sets. Nonparametric fit displayed for reference is the restricted-freedom monotone fit in Fig.\ \ref{fig:fits1} and \ref{fig:fits2}. The number of acceptable function variations sampled from the confidence set for each data realization is 5000.}
 \end{figure}

 Fig.\ \ref{fig:pd} shows the results of probing the 95\% confidence sets for uncertainties on peaks and dips,
 as outlined above, with 5000 acceptable function variations for each data realization.
 The box around a peak or a dip represents the largest horizontal and vertical variations in the scatter.
 In accordance with the confidence interval defined in Eq.\ \ref{eq:csetT},
 these form the 95\% confidence intervals on the location and height of a peak/dip.
 Table \ref{tab:pdci} lists these confidence intervals together with 95\% confidence intervals on peak height ratios.

 The following features of these results are worth pointing out.
 As is well-known, the first peak was very clearly resolved in the 1-year data itself.
 Our results are manifestly consistent with this observation,
 in the sense that its box is does not overlap with any other box.
 However, the uncertainty on the first peak does not shrink appreciably across the four data realizations.
 Further, our results clearly indicate that the second peak is resolved cleanly only in the 5-year data,
 whereas the third and fourth peaks are not resolved completely even in the 7-year data.

 \subsection{Uncertainties on the acoustic scale ($l_A$) and peak shift ($\phi_m$) parameters}

\begin{figure}
  \centerline{
  \begin{tabular}{c|c}
   \includegraphics[width=0.5\textwidth]{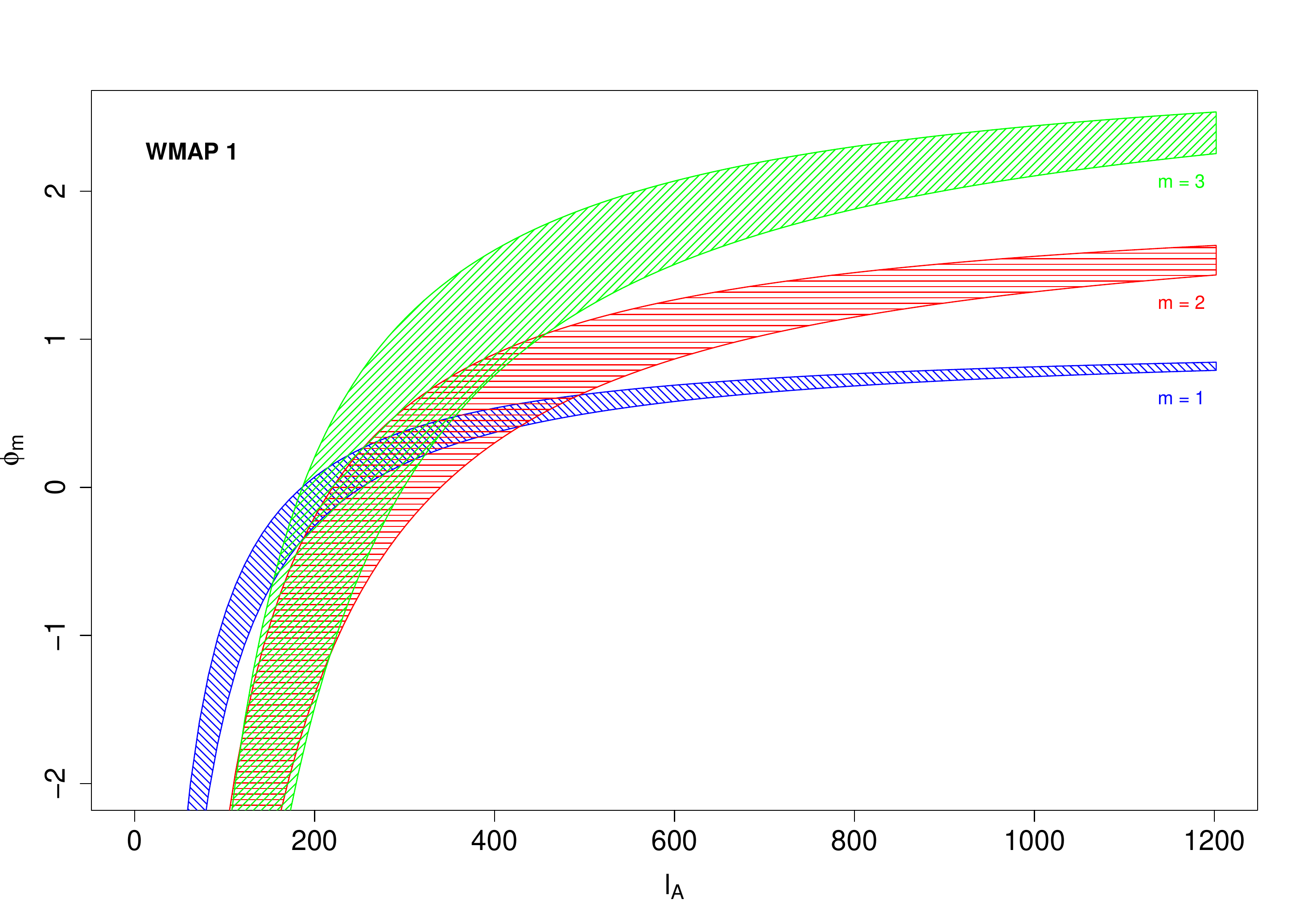}
   &
   \includegraphics[width=0.5\textwidth]{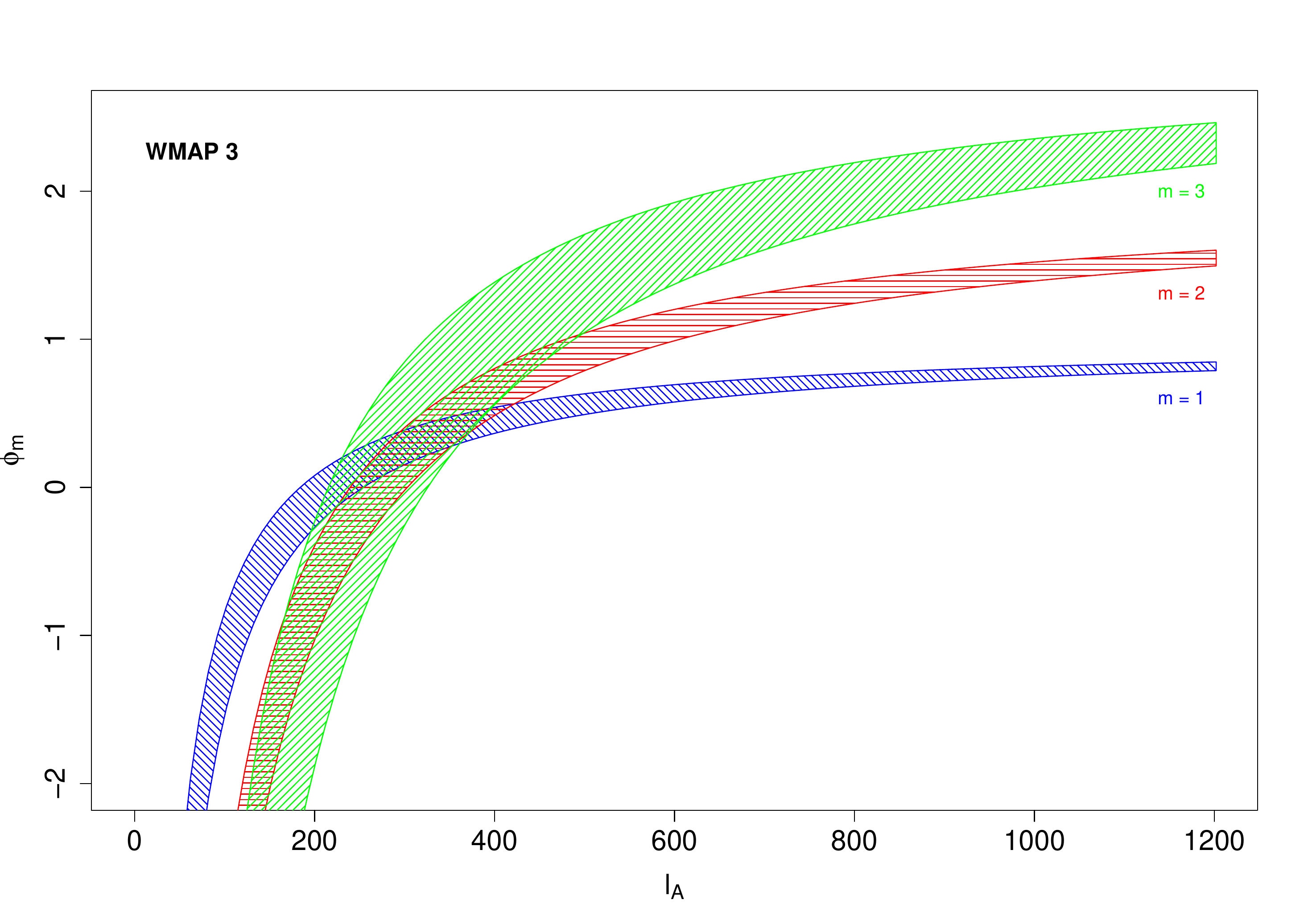}
   \\
   \hline
   \includegraphics[width=0.5\textwidth]{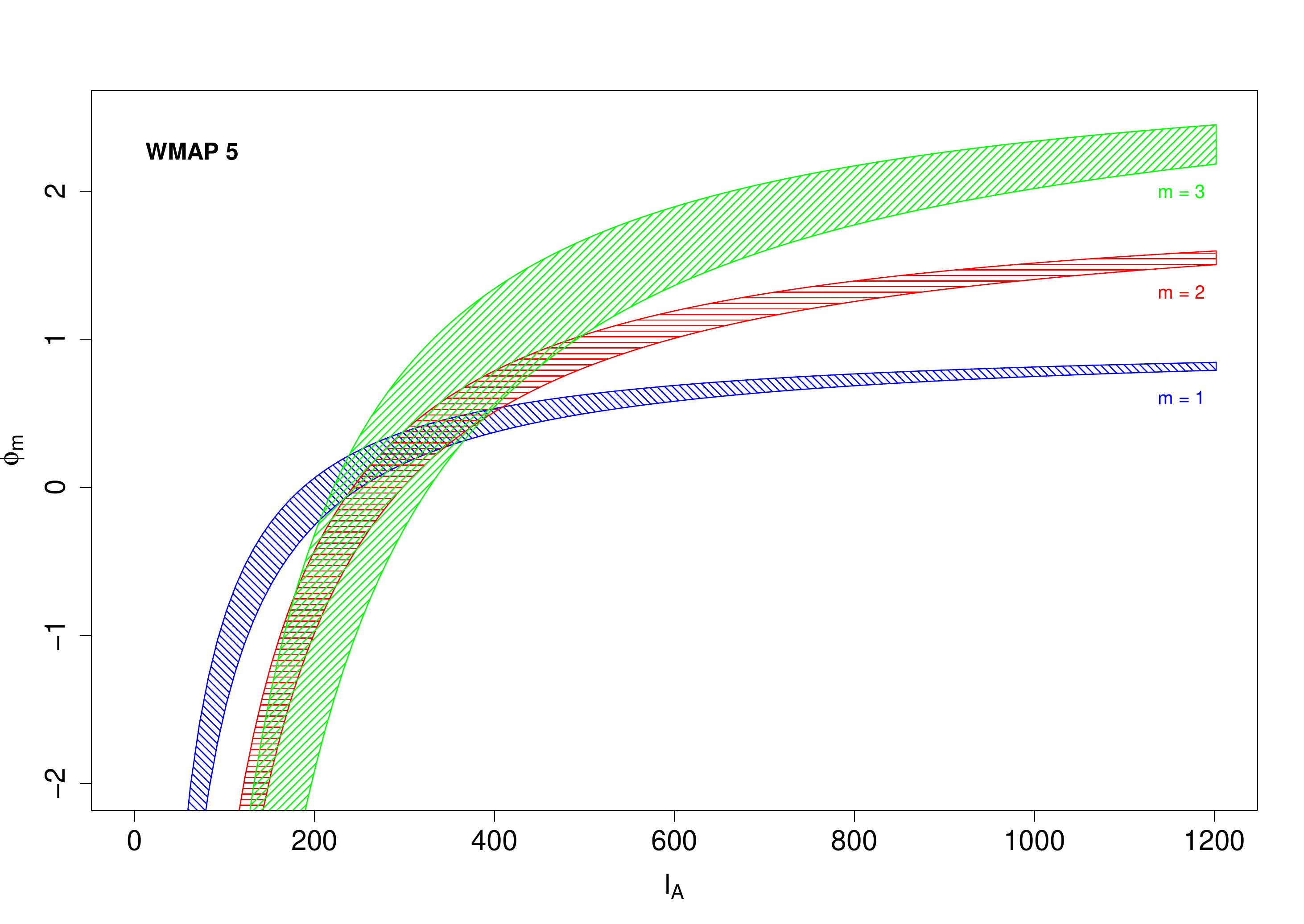}
   &
   \includegraphics[width=0.5\textwidth]{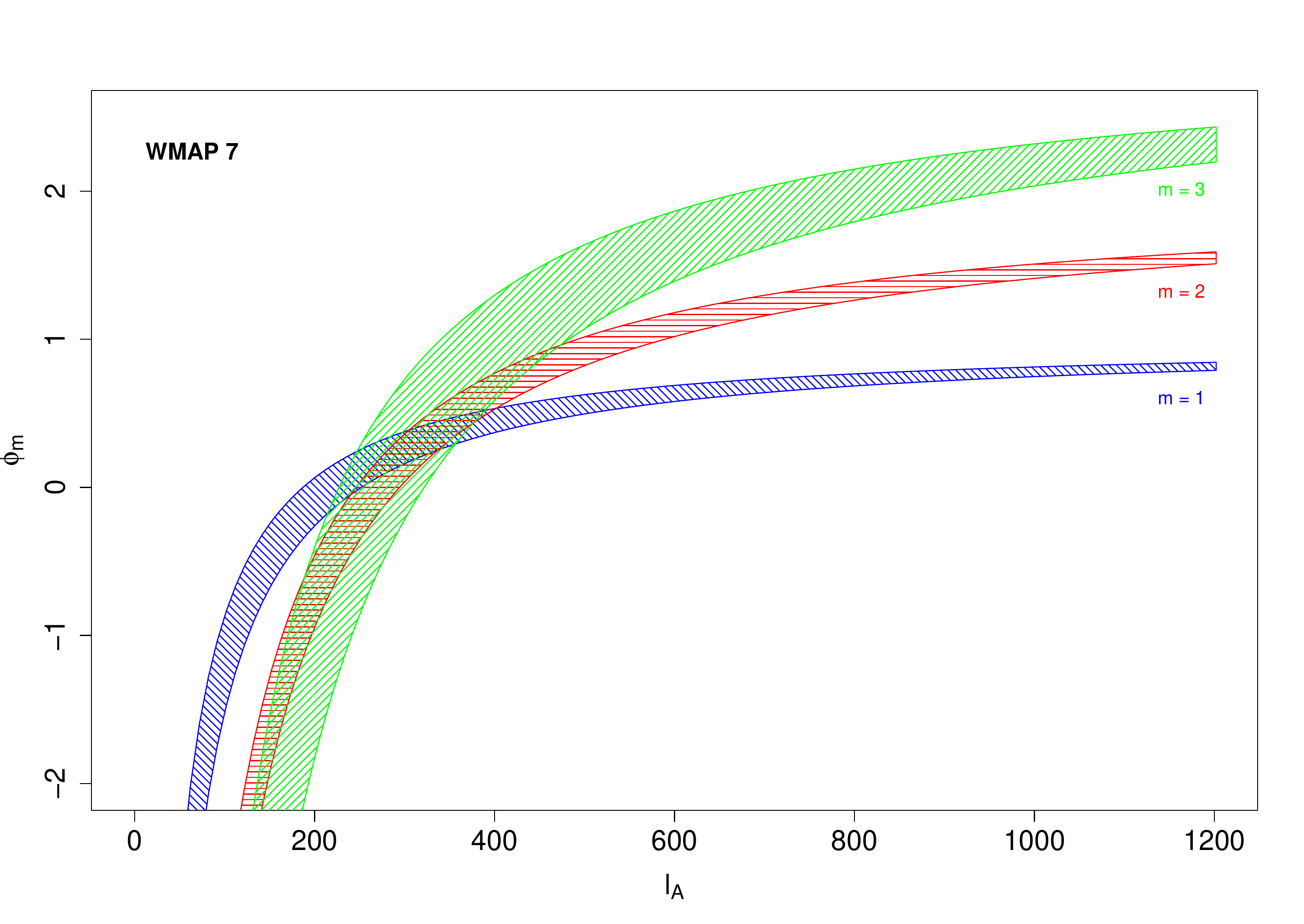}
  \end{tabular}
  }
  \caption{\label{fig:phi_vs_lA} Confidence ``bands'' for the acoustic scale $l_A$ and the shift $\phi_m$ for the $m$th peak, as derived from the 95\% confidence intervals on the first three peak locations (Table \ref{tab:pdci}) and Eq.\ \ref{eq:phi_lA}. Blue: $\phi_1$, red: $\phi_2$, green: $\phi_3$. Top left: WMAP 1-year, top right: 3-year, bottom left: 5-year, and bottom right: 7-year data sets. Note that these $(l_A,\phi_m)$ bands for different peaks $m$ appear to overlap around $\phi_m \approx 0$ and $ 200 \lesssim l_A \lesssim 400$: We interpret this as a nonparametric revelation of the nearly harmonic structure of peaks in the CMB power spectrum.}
 \end{figure}

 Consider the following relationship \citep{HFZ+2001,DL2002} between the location $l_m$ of the $m$th peak,
 the acoustic scale $l_A$, and the shift parameter $\phi_m$:
 \begin{equation}
  \label{eq:phi_lA}
  l_m = l_A ( m - \phi_m ).
 \end{equation}
 Substituting the end-points of the 95\% confidence interval for the $m$th peak location,
 this relationship results into a hyperbolic band of allowed values in the $l_A-\phi_m$ plane.
 Such bands, derived from 95\% confidence intervals on the first three peaks (Table \ref{tab:pdci}), are shown in Fig.\ \ref{fig:phi_vs_lA}.
 Additional information from other sources is required to constrain these bands to physically meaningful regions in the $l_A-\phi_m$ plane.
 For example, if we assume $l_A = 300$ \citep{PNB+2003}
 then, based on the 7-year data, the 95\% confidence intervals for $\phi_m$ will be $\phi_1: (0.1600,0.3767), \phi_2: (0.0367,0.3600), \phi_3: (-0.2167,0.7300)$.
 Conversely, additional constraints on $\phi_m$ could be used to generate a confidence interval on $l_A$.
 From a model-independent point of view, we note that the $(l_A,\phi_m)$ bands for different peaks $m$ appear to overlap around $\phi_m \approx 0$ and $ 200 \lesssim l_A \lesssim 400$.
 We interpret this as a nonparametric revelation of the nearly harmonic structure of peaks in the CMB power spectrum.

 \subsection{The low-$l$ up-turn from a nonparametric viewpoint}

 Another interesting feature in Fig.\ \ref{fig:pd} is the tiny
 but clearly observable scatter for the very first dip at the low-$l$ end.
 This scatter corresponds to extreme power spectrum variations that reside
 on the surface of the 95\% confidence set and have an up-turn at low $l$ values.
 In the $\Lambda$CDM cosmology, such up-turn at the low-$l$ end
 is primarily the result of the integrated Sachs-Wolfe (ISW) effect,
 and is seen in all parametric $\Lambda$CDM fits in Fig.\ \ref{fig:fits1} and \ref{fig:fits2}.
 It would therefore be interesting to see what could be said about the low-$l$ up-turn
 (and thereby about the ISW effect) based on the nonparametric confidence set.

 Notice that our nonparametric fits, which are at the center of their respective confidence sets,
 do not show a low-$l$ up-turn.
 However, the 7-year parametric $\Lambda$CDM fit, e.g., does show a clear up-turn at the low-$l$ end.
 This parametric fit is at a distance corresponding to confidence level of about 10\%
 ($\approx 0.12\sigma$; see Table \ref{tab:lambda}) from our 7-year nonparametric full-freedom monotone fit.
 This means that the confidence set for the 7-year nonparametric fit contains
 spectra with a low-$l$ up-turn at most as far away as the 7-year parametric fit.
 We therefore conclude conservatively that the low-$l$ up-turn as a feature of the CMB angular power spectrum
 cannot be ruled out at any confidence level in excess of about 10\%
 Actually, there are indications in our results (not shown) that such up-turned variations of the spectrum may be much closer to the center of the confidence set for the 7-year full-freedom monotone fit; this needs further investigation.

 \section{Conclusion}
 \label{conclusion}

 In this paper, we have presented a comparative nonparametric analysis of the
 WMAP 1-, 3-, 5-, and 7-year data releases for the CMB angular power spectrum,
 using a nonparametric function estimation methodology \citep{GMN+2004,BSM+2007}.
 In the context of this methodology, we have also presented our own numerical formulation
 for minimization of the inverse-noise-weighted risk function subject to monotonicity constraints,
 and a prescription for obtaining monotone nonparametric fits that are closer to cosmological expectations on smoothness.
 For all data realizations, we have presented results pertaining to the following questions:
 (a) how well is the angular power spectrum determined by the data alone,
 (b) how well is the $\Lambda$CDM model supported by a model-independent, nonparametric, data-driven analysis,
 and (c) what are the realistic uncertainties on peak/dip locations and heights.

 The motivation for the analysis presented here was to explore what could be inferred about
 the CMB angular power spectrum in a model-independent, data-driven manner.
 On the other hand, the basic physics of the CMB is quite well established.
 It would therefore be useful to connect a nonparametric/model-independent
 analysis such as ours with the known physics of the CMB angular power spectrum.
 This is reserved for the future.

 To conclude, we have demonstrated in this paper the threefold utility of
 the nonparametric methodology used here for cosmological function estimation problems:
 as a method with sound formal guarantees,
 as a sanity-enforcing mechanism on parametric model-based analyses,
 and as a method that allows interesting inferential questions to be
 addressed and answered in a data-driven manner.

 \acknowledgments
 MA is deeply indebted to Christopher R.\ Genovese and Larry Wasserman for many enlightening discussions covering all of statistics.
 Our \texttt{R} codes for computing the nonparametric fit are based on original codes by Christopher R.\ Genovese.
 TS would like to acknowledge support from the DST Swarnajayanti Fellowship.
 Insightful questions and suggestions from an anonymous referee helped improve
 the overall presentation aspects of the paper and, specifically, the method used in Sec.\ \ref{results:CoV}.

 \bibliographystyle{apj}
 \bibliography{cmb,cosmomc}

 \clearpage

 \begin{table}
   \caption{\label{tab:lambda}Distances of two model-based power spectra from our nonparametric fits}
   \begin{tabular}{lrr}
    \tableline\tableline

    Data              & $\Lambda$CDM & H$\Lambda$CDM \\

    \tableline\tableline

    1-year            &       0.1540 & 0.1493 \\
    $r_\alpha=0.3679$ &      16.70\% &  15.87\%  \\

    \tableline

    3-year            &       0.1462 & 0.2057 \\
    $r_\alpha=0.3653$ &      14.52\% &  29.03\%  \\

    \tableline

    5-year            &       0.1419 & 0.3552 \\
    $r_\alpha=0.3563$ &      19.66\% &  94.82\%  \\

    \tableline

    7-year            &       0.1238 & 0.3550 \\
    $r_\alpha=0.3551$ &       9.08\% &  94.98\%  \\

    \tableline
  \end{tabular} 
  \tablecomments{These are the distances of two model-based ($\Lambda$CDM and H$\Lambda$CDM) power spectra from our nonparametric (full-freedom monotone) fits. The H$\Lambda$CDM model \citep{PHK+1995,PG2001} considered here for illustrative purposes is defined by a small neutrino fraction ($\Omega_\nu h^2 = 0.00275$) with corresponding adjustment to the dark energy content ($\Omega_\Lambda=0.729756$), and the rest of the parameters (including zero curvature) being identical to that of the best $\Lambda$CDM model \citep{LDH+2011} for the 7-year data (power spectrum generated using the CAMB software \citep{Lewis:1999bs}). $\Lambda$CDM-based fits used are the best parametric fits for the corresponding data realization \citep{HSV+2003,HNB+2007,NDH+2009,LDH+2011}. $r_\alpha$ is the confidence radius at $\alpha = 0.05$ (i.e., 95\% confidence level $\equiv 2 \sigma$). Percentages reported are the confidence levels corresponding to these distances; these can be interpreted as asymptotic probabilities with which the corresponding parametric fit is ruled out as a candidate for the true but unknown spectrum. Notice the dramatic progression, as the data become precise, of (a) how this H$\Lambda$CDM model is pushed to the boundary of the confidence set, and (b) how the $\Lambda$CDM model gets closer to the nonparametric fit.}
 \end{table}

 \clearpage

 \begin{deluxetable}{lrrrrcc}
  \tabletypesize{\scriptsize}
  \rotate
  \tablecaption{\label{tab:pdci}95\% confidence intervals on several features of the angular power spectrum}
  \tablewidth{0pt}
  \tablehead{   
  \colhead{Data} & \colhead{Peak Location} &  \colhead{Peak Height} & \colhead{Dip Location} & \colhead{Dip Height} & \colhead{Peak Location Ratio} & \colhead{Peak Height Ratio}
}
\startdata
   1-year & $l_1: \hfill  (186,252)$ & $h_1: \hfill (4968, 6133)$ & $l_{1+{1 \over 2}}: \hfill  (378,507)$ & $h_{1+{1 \over 2}}: \hfill  ( 1152, 2029) $ & ... & ... \\
          & $l_2: \hfill  (440,680)$ & $h_2: \hfill (1879, 3933)$ & $l_{2+{1 \over 2}}: \hfill  (494,772)$ & $h_{2+{1 \over 2}}: \hfill  (-2929, 2868) $ & $l_2/l_1: \hfill (1.953,3.217)$ & $h_2/h_1: \hfill (0.329,0.724)$ \\
          & $l_3: \hfill  (559,897)$ & $h_3: \hfill (41,  10733)$ & $l_{3+{1 \over 2}}: \hfill  (639,900)$ & $h_{3+{1 \over 2}}: \hfill  (-11376,9962) $ & $l_3/l_1: \hfill (2.495,4.385)$ & $h_3/h_1: \hfill  (0.0075,1.906)$ \\
\tableline
   3-year & $l_1: \hfill  (184,254)$ & $h_1: \hfill (5130, 6199)$ & $l_{1+{1 \over 2}}: \hfill  (384,456)$ & $h_{1+{1 \over 2}}: \hfill  ( 1402, 1917) $ & ... & ... \\
          & $l_2: \hfill  (479,606)$ & $h_2: \hfill (2172, 2997)$ & $l_{2+{1 \over 2}}: \hfill  (593,856)$ & $h_{2+{1 \over 2}}: \hfill  ( 80, 2145) $ & $l_2/l_1: \hfill  (2.028,3.000)$ & $h_2/h_1: \hfill (0.380,0.555)$ \\
          & $l_3: \hfill  (646,977)$ & $h_3: \hfill (1506, 6045)$ & $l_{3+{1 \over 2}}: \hfill (729,1000)$ & $h_{3+{1 \over 2}}: \hfill  (-4501, 5113) $ & $l_3/l_1: \hfill  (2.761,4.801)$ & $h_3/h_1: \hfill  (0.264,1.082)$ \\
\tableline
   5-year & $l_1: \hfill  (187,251)$ & $h_1: \hfill (5249, 6301)$ & $l_{1+{1 \over 2}}: \hfill  (387,445)$ & $h_{1+{1 \over 2}}: \hfill  ( 1489, 1934) $ & ... & ... \\
          & $l_2: \hfill  (485,596)$ & $h_2: \hfill (2306, 2955)$ & $l_{2+{1 \over 2}}: \hfill  (608,793)$ & $h_{2+{1 \over 2}}: \hfill  ( 971, 2095) $ & $l_2/l_1: \hfill  (2.040,2.963)$ & $h_2/h_1: \hfill (0.381,0.525)$ \\
          & $l_3: \hfill  (663,982)$ & $h_3: \hfill (1863, 4635)$ & $l_{3+{1 \over 2}}: \hfill (731,1000)$ & $h_{3+{1 \over 2}}: \hfill  (-2978, 3157) $ & $l_3/l_1: \hfill  (2.883,4.672)$ & $h_3/h_1: \hfill (0.321,0.833)$ \\
\tableline
   7-year & $l_1: \hfill  (187,252)$ & $h_1: \hfill (5177, 6377)$ & $l_{1+{1 \over 2}}: \hfill  (390,442)$ & $h_{1+{1 \over 2}}: \hfill  (1512,1931) $ & ... & ... \\
          & $l_2: \hfill  (492,589)$ & $h_2: \hfill (2328, 3015)$ & $l_{2+{1 \over 2}}: \hfill  (623,803)$ & $h_{2+{1 \over 2}}: \hfill  (1074,2063) $ & $l_2/l_1: \hfill (2.060,2.887)$ & $h_2/h_1: \hfill ( 0.386, 0.534)$ \\
          & $l_3: \hfill  (681,965)$ & $h_3: \hfill (1871, 4119)$ & $l_{3+{1 \over 2}}: \hfill (733,1104)$ & $h_{3+{1 \over 2}}: \hfill  (-3111, 2709) $ & $l_3/l_1: \hfill (3.000,4.553)$ & $h_3/h_1: \hfill ( 0.323, 0.732)$ \\
          & $l_4: \hfill (815,1193)$ & $h_4: \hfill (-1391,7726)$ & $l_{4+{1 \over 2}}: \hfill (951,1200)$ & $h_{4+{1 \over 2}}: \hfill  (-11102,6364) $ & $l_4/l_1: \hfill (3.606,6.047)$ & $h_4/h_1: \hfill (-0.244, 1.340)$ \\
  \enddata
  \tablecomments{Negative values for heights and height ratios is a reflection of the high noise at the high-$l$ end. Here, $l_m$ ($h_m$) stands for the location (height) of the $m$th peak, and $l_{m+{1 \over 2}}$ ($h_{m+{1 \over 2}}$) denotes the location (depth) of the $m$th dip.}
 \end{deluxetable}
\end{document}